\documentclass[twocolumn,english]{article}
\usepackage{ae,aecompl}
\usepackage[T1]{fontenc}
\usepackage[latin9]{inputenc}
\usepackage[a4paper]{geometry}
\usepackage{babel}
\usepackage{bm}
\usepackage{amsmath}
\usepackage{amssymb}
\usepackage{graphicx}
\usepackage[unicode=true]
 {hyperref}

\makeatletter

\providecommand{\tabularnewline}{\\}
\newcommand{\lyxdot}{.}

\newcommand{\lyxaddress}[1]{
	\par {\raggedright #1
	\vspace{1.4em}
	\noindent\par}
}

\usepackage{babel}
\usepackage{xcolor}
\definecolor{myred}{rgb}{0.66, 0.15, 0.15}
\definecolor{darkgreen}{rgb}{0.0, 0.5, 0.0}

\usepackage{hyperref}          
\hypersetup{
     colorlinks   = true,
     citecolor    = myred,
     urlcolor     = myred,
     urlbordercolor = myred
}

\newcommand{\rs}{\bar{r}_e}
\newcommand{\alphas}{~\bar{\alpha}}
\newcommand{\rsname}{effective range at resonance }
\newcommand{\alphasname}{width parameter}
\usepackage[normalem]{ulem}
\usepackage{flushend}
\usepackage{cuted}

\usepackage{cite}

\makeatother

\begin{document}
\begin{strip}\textsf{\textbf{\huge{}Shallow Trimers of Two Identical
Fermions and One Particle in Resonant Regimes}}{\huge\par}

{\Large{}\vspace{0.8cm}
Pascal Naidon$^{1}$, Ludovic Pricoupenko$^{2}$, Christiane Schmickler$^{1}$}
{\Large{}\vspace{0.5cm}
 }\today\\
 \vspace{0.5cm}

\lyxaddress{{\small{}{}$^{1}$RIKEN Nishina Centre, Strangeness Nuclear Physics
Laboratory, RIKEN, Wak{\={o}}, 351-0198 Japan.}\\
 \textit{\href{mailto:christiane.schmickler@riken.jp}{christiane.schmickler@riken.jp},
\href{mailto:pascal@riken.jp}{pascal@riken.jp}}}

\lyxaddress{{\small{}{}$^{2}$Sorbonne Universit\'e, CNRS, Laboratoire de Physique
Th\'eorique de la Mati\`ere Condens\'ee (LPTMC), F-75005 Paris,
France.}\\
 \textit{\href{mailto:ludovic.pricoupenko@sorbonne-universite.fr}{ludovic.pricoupenko@sorbonne-universite.fr}}}
\begin{abstract}
We consider two identical fermions interacting in the p-wave channel.
Each fermion also interacts with another particle in the vicinity
of an s-wave resonance. We find that in addition to the Kartavtsev-Malykh
universal trimer states resulting from the s-wave particle-fermion
interaction, the fermion-fermion p-wave interaction induces one or
two shallow trimers in a large domain of the control parameters, including
a borromean regime where the ground-state trimer exists in the absence
of dimers at any mass ratio between the fermions and the particle.
A generic picture of the trimer spectrum emerges from this work in
terms of the low-energy parameters of the interactions. 
\end{abstract}
\end{strip}

\section{Introduction}

\hypersetup{linkcolor=myred}

The experimental achievement of Fano-Feshbach resonances in ultra-cold
atomic systems~\cite{Inouye1998} have enabled researchers to modify
interatomic interactions at will. In particular, it is possible to
tune the scattering length of the atoms to values much larger than
the interaction range and thus to achieve quantum systems of particles
with short-range resonant interactions, whose properties are universally
described by only a few parameters. At the two-body level, two atoms
undergoing such a resonant interaction in the s-wave can form halo
dimers, whose binding energy is universally given by the square inverse
of the scattering length. At the three-body level, it is possible
to form Efimov trimer states~\cite{Efimov1970a}, which are universally
described by the scattering length and a three-body parameter~\cite{Braaten2006,Naidon2017,Greene2017}.
The most striking property of these states is that they exhibit the
Efimov effect: at resonance, where a halo dimer becomes bound, the
trimer spectrum forms an infinite geometric series with an accumulation
point at vanishing energy. This effect results from the Efimov attraction,
an effective three-body attraction due to the resonant nature of the
interaction and which arises far outside the range of inter-particle
interactions. The experimental confirmation of Efimov trimers~\cite{Kraemer2006,Ottenstein2008,Knoop2009,Zaccanti2009,Pollack2009,Gross2009,Wenz2009,Huckans2009,Williams2009,Lompe2010a,Gross2010,Nakajima2010,Lompe2010,Gross2011,Nakajima2011,Machtey2012,Knoop2012,Roy2013,Dyke2013,Huang2014,Tung2014,Pires2014}
with ultracold atoms has spurred an intense theoretical search for
universal few-body states in systems with resonant interactions.

The Efimov effect is most easily evidenced in bosonic systems, and
tends to be absent in fermionic systems, which involve non-zero angular
momenta due to the Pauli exclusion. Indeed, a non-zero angular momentum
implies a centrifugal repulsion, which often overcomes the Efimov
attraction. For instance, three-body systems of spin-1/2 fermions
do not exhibit the Efimov effect for this reason. The relative strength
of the Efimov attraction with respect to the centrifugal repulsion
can nonetheless be enhanced in the case of particles of different
masses. The simplest system consists of two identical fermions of
mass $M$ in the same spin state interacting with a third particle
(fermion or boson) of mass $m$. The third particle experiences an
s-wave scattering resonance with each fermion. Analogously to the
exchange of mesons between two nucleons in nuclear physics, it mediates
an effective attraction between the two fermions. The fact that the
two identical fermions are exactly in the same internal state implies
that they have at least one unit of relative angular momentum, resulting
in a centrifugal repulsion competing with the Efimov attraction. When
the mass ratio $x=M/m$ exceeds a critical value $x_{c}=13.607$ \dots ,
the Efimov attraction dominates over the centrifugal repulsion and
the Efimov effect occurs at resonance, resulting in an infinite number
of trimers with the $J^{\pi}=1^{-}$ symmetry. At finite but large
and positive scattering length, when the mass ratio is smaller than
this critical value, the effective three-body attraction persists
at intermediate distances, making it possible for the three particles
to bind into a finite number of bound states.

Considering two fermions interacting with the third particle only
through a contact interaction, Kartavtsev and Malykh have demonstrated
that for mass ratios larger than $x_{1}=8.17260...$ and smaller than
$x_{c}$, up to three trimer states can exist~\cite{Kartavtsev2007,Kartavtsev2008,Kartavtsev2009,Kartavtsev2016}.
These trimers have the same $J^{\pi}=1^{-}$ symmetry as Efimov trimers
at larger mass ratios and are characterised by the scattering length
$a$ between each of the two fermions and the third particle, and
additionally, for $M/m$ larger than $x_{r}=8.619\dots$, by a three-body
parameter. These universal results are applicable to describe shallow
trimer states in real systems in the limit of large scattering length
$a$ and negligible interactions between the fermions. There have
been some attempts~\cite{Endo2012,Safavi-Naini2013,Castin2011} to
understand more precisely how this universal scenario fits into real
systems. In what follows, we will call \emph{s-wave induced trimers}
the states bound only by the s-wave interaction between the fermions
and particle, and use the short-hand notation \emph{KM states} or
\emph{KM limit} to refer to their universal limit described by Kartavtsev
and Malykh's contact theory.

One pending question is how the universal scenario is modified by
the presence of an interaction between the two fermions. In real systems,
an interaction is indeed always present. If sufficiently attractive,
in the vicinity of a two-body p-wave scattering resonance, this interaction
may even bind the two fermions, despite their centrifugal repulsion.
This attractive effect is described at low energy by only two parameters,
the p-wave scattering volume and the p-wave effective range. One may
thus wonder whether the universality of the KM states is preserved
in the presence of the fermions' interaction, i.e. whether the system
can still be universally described by a few low-energy parameters.

\begin{figure}
\includegraphics[width=8cm]{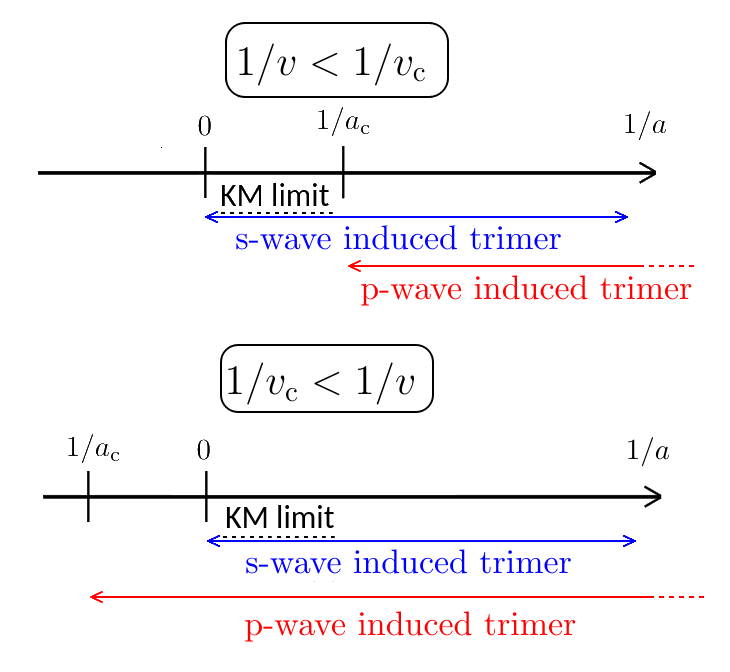}

\caption{\label{fig:Schematic-representation}Domains of existence of the two
types of trimers identified in this work, for a mass ratio $M/m\in[x_{1}\approx8.1726,x_{c}^{\prime}<x_{2}]$.
The \emph{s-wave induced trimer}, shown schematically in blue, is
induced by the s-wave interaction between the fermions and the particle.
It exists only for positive s-wave scattering lengths $a>0$ and admits
a universal limit (KM limit) for $1/a\to0^{+}$. The \emph{p-wave
induced trimer}, shown in red, is induced by the p-wave interaction
between the fermions in presence of the s-wave interaction. It exists
only for positive $1/a>1/a_{c}$ when the p-wave inverse scattering
volume $1/v$ is less than a critical value $1/v_{c}$ (upper panel),
and can exist for negative $1/a$ (Borromean trimer) when $1/v>1/v_{c}$
(lower panel). Not shown in the figure: for larger mass ratios $M/m>x_{2}$,
there is a second s-wave induced trimer, converging to the excited
KM state for $1/a\to0^{+}$. For mass ratios $M/m>x_{c}^{\prime},$there
is an excited p-wave induced trimer. Finally, for $M/m>x_{c}\approx13.607$,
there are no more KM states for $1/a\to0^{+}$ but an infinite number
of s-wave induced trimers converging to Efimov states.}

\end{figure}

In this paper, we show that the universal KM states always exist for
the mass ratios predicted by the contact theory and for a sufficiently
large and positive value of the scattering length. However, for sufficiently
attractive p-wave interaction, the spectrum is enriched by another
shallow state whatever the mass ratio, and by a second one for a mass
ratio $M/m$ larger than a critical value $x_{c}^{\prime}$. Hereafter,
we will use the denomination \emph{p-wave induced trimers} for these
states. Their properties (threshold, energy\dots ) depend on the shape
of the p-wave interaction. Despite this non-universality, these results
permit us to draw a generic picture for the spectrum of the shallow
trimers in terms of s-wave and p-wave induced trimers, as a function
of the two-body low-energy parameters. Both the s-wave and p-wave
induced trimers are depicted schematically in Fig.~\ref{fig:Schematic-representation}
as a function of the s-wave scattering length $a$ and the p-wave
scattering volume $v$. As will be discussed, these two types of trimers
may undergo an avoided crossing and hybridise when their energies
come close.

\section{Model}

\subsection{Separable model}

We consider the general problem of two identical fermions of mass
$M$ and one particle of mass $m$, all three interacting with each
other through finite-range attractive interactions. The two fermions
are denoted as particles 1 and 2, and the third particle as particle
3. In these notations, the Schrödinger equation at energy $E$ reads
in momentum representation: 
\begin{equation}
\left(\frac{\hat{p}_{1}^{2}}{2M}+\frac{\hat{p}_{2}^{2}}{2M}+\frac{\hat{p}_{3}^{2}}{2m}+\hat{U}_{12}+\hat{V}_{23}+\hat{V}_{31}-E\right)\vert\Psi\rangle=0\label{eq:SchrodingerEq}
\end{equation}
where $\hat{p}_{i}$ is the momentum operator of particle $i$, and
$\hat{V}_{ij}$ and $\hat{U}_{ij}$ are the interaction operators
between particles $i$ and $j$. No three-body interaction is considered
in this work. The state $\vert\Psi\rangle$ is antisymmetric under
the exchange of fermions, i.e. $\langle\bm{k}_{2},\bm{k}_{1},\bm{k}_{3}\vert\Psi\rangle=-\langle\bm{k}_{1},\bm{k}_{2},\bm{k}_{3}\vert\Psi\rangle$
in momentum representation where $\bm{k}_{i}$ is the wave vector
of particle $i$. Because of translational invariance, we can pursue
this study in the center-of-mass frame and thus $\bm{k}_{1}+\bm{k}_{2}+\bm{k}_{3}=\bm{0}$.
The wave function can then be expressed in terms of one of the three
Jacobi coordinate sets, $(\bm{k}_{i},\bm{k}_{jk})$, where $\bm{k}_{jk}\equiv\frac{m_{j}\bm{k}_{k}-m_{k}\bm{k}_{j}}{m_{j}+m_{k}}$
is the relative momentum between particles $j$ and $k$, with $m_{1}=m_{2}=M$
and $m_{3}=m$.

To simplify the calculations, the interaction operators are taken
to be of the separable type: 
\begin{align}
\hat{V} & =\frac{2\pi\hbar^{2}}{\mu_{23}}\xi\vert\chi\rangle\langle\chi\vert\label{eq:V}\\
\hat{U} & =\frac{6\pi\hbar^{2}}{\mu_{12}}\sum_{m=-1}^{1}g_{m}\vert\Phi_{m}\rangle\langle\Phi_{m}\vert\label{eq:U}
\end{align}
where $\mu_{23}=\left(\frac{1}{M}+\frac{1}{m}\right)^{-1}$ and $\mu_{12}=M/2$
are the reduced masses of the fermion-particle and two-fermion subsystems.
The separable potentials are characterised by interaction strengths
$\xi\le0$ and $g_{m}\le0$, and form factors $\chi$ and $\Phi_{m}$,
with $\langle\bm{k}\vert\chi\rangle\equiv\chi(k)$ and $\langle\bm{k}\vert\Phi_{m}\rangle\equiv\Phi_{m}(\bm{k})=\phi_{m}(k)\bm{k}\cdot\hat{\bm{e}}_{m}$,
where we use the unit vectors $\hat{\bm{e}}_{0}=\hat{\bm{e}}_{z}$
and $\hat{\bm{e}}_{\pm1}=(\mp\hat{\bm{e}}_{x}-i\hat{\bm{e}}_{y})/\sqrt{2}$.
The form of these separable potentials is chosen such that $\hat{V}$
only affects the s wave, and $\hat{U}$ only affects the p wave -
see Appendix~1. In general the interaction $\hat{U}$ can be anisotropic,
and becomes isotropic in the case where all the interaction strengths
$g_{m}$ and form factors $\phi_{m}$ are equal. In this paper, without
loss of generality we will choose the normalisation of the form factors
such that $\phi_{m}(0)=\chi(0)=1$.

Thanks to the separability of the potentials, the problem can be formulated
in terms of the quantities, 
\begin{align}
S(\bm{k}_{1}) & =-\xi\int\frac{d^{3}k_{23}}{(2\pi)^{3}}\langle\chi\vert\bm{k}_{23}\rangle\langle\{\bm{k}_{i}\}\vert\Psi\rangle\label{eq:S}\\
P_{m}(\bm{k}_{3}) & =-g_{m}\int\frac{d^{3}k_{12}}{(2\pi)^{3}}\langle\Phi_{m}\vert\bm{k}_{12}\rangle\langle\{\bm{k}_{i}\}\vert\Psi\rangle.\label{eq:P}
\end{align}
Physically, $S$ corresponds to the relative wave function between
a fermion-particle pair and the other fermion, and $P_{m}$ to the
relative wave function between the two-fermion pair and the third
particle when the fermion pair has an angular momentum projection
$m\hbar$. They satisfy the generalised Skorniakov--Ter-Martirosian
(STM) equations (see appendix~2): 
\begin{align}
-\frac{\vert\chi(i\kappa_{0})\vert^{2}}{4\pi f_{0,0}(i\kappa_{0})}S(\bm{k})=\qquad\qquad\qquad\qquad\qquad\qquad\label{eq:STM1}\\
y\int\frac{d^{3}k^{\prime}}{(2\pi)^{3}}\frac{\langle\chi\vert\bm{k}^{\prime}+\frac{x}{2y}\bm{k}\rangle\langle\bm{k}+\frac{x}{2y}\bm{k}^{\prime}\vert\chi\rangle S(\bm{k}^{\prime})}{y(k^{2}+k^{\prime2})+x\bm{k}\cdot\bm{k}^{\prime}+q^{2}}\nonumber \\
+3\int\frac{d^{3}k^{\prime}}{(2\pi)^{3}}\frac{\langle\chi\vert\bm{k}^{\prime}+\frac{1}{2y}\bm{k}\rangle\sum_{m}\langle\bm{k}+\frac{1}{2}\bm{k}^{\prime}\vert\Phi_{m}\rangle P_{m}(\bm{k}^{\prime})}{k^{2}+yk^{\prime2}+\bm{k}\cdot\bm{k}^{\prime}+q^{2}}\nonumber 
\end{align}
\begin{multline}
\frac{\vert\kappa_{1}\phi_{m}(i\kappa_{1})\vert^{2}}{4\pi f_{1,m}(i\kappa_{1})}P_{m}(\bm{k})=\\
2y\int\frac{d^{3}k^{\prime}}{(2\pi)^{3}}\frac{\langle\Phi_{m}\vert\bm{k}^{\prime}+\frac{\bm{k}}{2}\rangle\langle\bm{k}+\frac{\bm{k}^{\prime}}{2y}\vert\chi\rangle S(\bm{k}^{\prime})}{yk^{2}+k^{\prime2}+\bm{k}\cdot\bm{k}^{\prime}+q^{2}}\label{eq:STM2}
\end{multline}
where we have introduced the binding wave number $q$ of the system
defined by 
\begin{equation}
E=-\frac{\hbar^{2}q^{2}}{M},\label{eq:Energy}
\end{equation}
as well as the following short-hand notations for numerical factors
depending only on the mass ratio,
\begin{align}
x & \equiv M/m\label{eq:x}\\
y & \equiv(x+1)/2\label{eq:y}
\end{align}
There are two kinds of terms in the STM equations: the terms on the
right-hand side, which contain integrals describing the three-body
sector, and the terms on the left-hand side, which describe only the
two-body sector and contain the s-wave and p-wave two-body scattering
amplitudes $f_{0,0}$ and $f_{1,m}$ given by 
\begin{align}
-\frac{\vert\chi(i\kappa)\vert^{2}}{f_{0,0}(i\kappa)} & =\frac{1}{\xi}+\frac{2}{\pi}\int_{0}^{\infty}dk\,k^{2}\frac{\vert\chi(k)\vert^{2}}{k^{2}+\kappa^{2}}\label{eq:f0}\\
\frac{\vert\kappa\phi_{m}(i\kappa)\vert^{2}}{f_{1,m}(i\kappa)} & =\frac{1}{g_{m}}+\frac{2}{\pi}\int_{0}^{\infty}dk\,k^{2}\frac{\vert k\phi_{m}(k)\vert^{2}}{k^{2}+\kappa^{2}}\label{eq:f1}
\end{align}
These amplitudes are evaluated at the relative wave number $i\kappa$
between two particles in the presence of a third particle of wave
number $k$ with respect to that pair, for a fixed three-body energy
$E$. This binding wave number is given, for the s-wave and p-wave
respectively, by $\kappa_{0}$ and $\kappa_{1}$ satisfying
\begin{align}
\kappa_{0}^{2} & \equiv\frac{2x+1}{4y^{2}}k^{2}+q^{2}/y,\label{eq:kappas}\\
\kappa_{1}^{2} & \equiv\frac{2x+1}{4}k^{2}+q^{2}.\label{eq:kappap}
\end{align}

In what follows, we will introduce the scale $\Lambda_{0}$ (resp.
$\Lambda_{1}$) below which the form factor $\chi(k)$ (resp. $\phi(k)$)
are almost constant (i.e. equal to unity in our choice of normalisation).
Physically these scales are related to the radius of the actual interactions
which are supposed to be of short range. In the context of ultracold
atoms they are of the order of the inverse of the van der Waals lengths
$\Lambda_{0}\propto\left(\frac{\hbar^{2}}{2\mu_{23}C_{23}}\right)^{1/4}$
and $\Lambda_{1}\propto\left(\frac{\hbar^{2}}{2\mu_{12}}C_{12}\right)^{1/4}$,
where $C_{23}$ and $C_{12}$ are the dispersion coefficients of the
$-1/r^{6}$ van der Waals term in the respective pairwise potential
for the pairs of particles $(12)$ and $(23)$. In the small momentum
limit where $\kappa$ is much smaller than $\Lambda_{0}$ and $\Lambda_{1}$,
one has the usual expansions of the scattering amplitudes: 
\begin{align}
\frac{1}{f_{0,0}(i\kappa)} & =-\frac{1}{a}+\kappa-\frac{r_{{\rm e}}}{2}\kappa^{2}+o(\kappa^{2})\label{eq:swaveAmplitude_Expansion}\\
\frac{-\kappa^{2}}{f_{1,m}(i\kappa)} & =-\frac{1}{v_{m}}+\alpha_{m}\kappa^{2}-\kappa^{3}+o(\kappa^{3})\label{eq:pwaveAmplitude_Expansion}
\end{align}
where $a$ is the s-wave scattering length and $v_{m}$ is the scattering
volume, which in general depends on the quantum number $m$ of the
fermionic pair. Physical short-range potentials admit the same expansions
as Eqs.~(\ref{eq:swaveAmplitude_Expansion}-\ref{eq:pwaveAmplitude_Expansion}).
At the two-body level, the separable potentials are therefore indistinguishable
from physical ones in this low-momentum limit, which motivates their
choice for the study of low-energy physics. However, the separability
induces discrepancies at higher momenta comparable or larger than
$\Lambda_{0},\Lambda_{1}$, which may affect the properties of low-energy
three-body states. For instance, the three-body parameter of Efimov
states may differ, although these differences are generally small
as long as the scale $\Lambda_{0}^{-1}$ is properly set to the physical
range of interactions. Separable s-wave potentials have been shown
to be a good approximation for shallow potentials supporting at most
one dimer \cite{Naidon2014,Mestrom2019}. We assume that in a similar
fashion, the p-wave separable potential is an adequate approximation
for a qualitative description of low-energy three-body states, as
long as the scale $\Lambda_{1}^{-1}$ corresponds to the physical
range of interaction.

By taking the limit $\kappa\to0$ of Eqs.~(\ref{eq:f0}-\ref{eq:f1}),
one finds the relations between the interaction strengths $\xi$ and
$g_{m}$ and the scattering length and volume $a$ and $v_{m}$: 
\begin{align}
\frac{1}{a} & =\frac{1}{\xi}+\frac{2}{\pi}\int_{0}^{\infty}dk\vert\chi(k)\vert^{2}\label{eq:inversea}\\
\frac{1}{v_{m}} & =\frac{1}{g_{m}}+\frac{2}{\pi}\int_{0}^{\infty}dkk^{2}\vert\phi_{m}(k)\vert^{2}.\label{eq:inversev}
\end{align}

The coefficients $r_{{\rm e}}$ and $1/\alpha_{m}$ in Eqs.~(\ref{eq:swaveAmplitude_Expansion},\ref{eq:pwaveAmplitude_Expansion})
are the s-wave and p-wave effective ranges, which in general depend
on $a$ and $v_{m}$, respectively. Although the effective ranges
are useful to describe the low-energy two-body physics, as their name
implies, they do not represent the true range of interactions. For
near-resonant three-body systems, it is more useful to consider the
value $\rs$ of the effective range $r_{e}$ at the s-wave resonance
($1/a\to0$), and the value $1/\alphas_{m}$ of the effective range
$1/\alpha_{m}$ at the p-wave resonance ($1/v_{m}\to0$). For the
single-channel interactions considered in the present study, $\rs$
and $1/\alphas_{m}$ are of the order of the potential radii $1/\Lambda_{0}$
and $1/\Lambda_{1}$, and have, unlike $\Lambda_{0}$ and $\Lambda_{1}$,
a precise definition from Eqs.~(\ref{eq:swaveAmplitude_Expansion},\ref{eq:pwaveAmplitude_Expansion}).
For this reason, we will compare different interaction models having
the same scattering lengths/volumes, and the same effective ranges
in the limits $1/a\to0$ or $1/v_{m}\to0$. This ensures that these
models have the same two-body spectra near these limits and also have
the same interaction ranges. In this spirit, the s-wave \rsname ${\rs}$
will be used as the unit of length throughout this paper.

\subsection{Isotropic p-wave interaction \label{sec:symmetry}}

The isotropic p-wave interaction where $g_{m}=g$ and $\phi_{m}=\phi$
plays a central role in the subsequent analysis. In this section,
we thus simplify the STM equations by using symmetry considerations
in this particular case. Moreover, we will show that the results in
the isotropic limit can be directly used for a more realistic anisotropic
p-wave interaction by using a perturbative treatment. In what follows,
we will use the notation $f_{1,m}\equiv f_{1}$, $v_{m}\equiv v$
and $\alphas_{m}=\alphas$. Before going to the three-body equations,
it is worth recalling some important and general properties of isotropic
p-wave interactions in the resonant regime \cite{Landau_meca_Q,Pricoupenko2006,Pricoupenko2006a}: 
\begin{enumerate}
\item[$i)$] the scattering resonance due to a quasi-bound state occurs for large
and negative values of the scattering volume ($1/v\to0^{-}$) at a
specific value of the relative momentum $k_{{\rm res}}=1/\sqrt{-\alphas v}$; 
\item[$ii)$] the width of the scattering resonance is of the order of $k_{{\rm res}}/\alphas$,
which is why the inverse p-wave effective range at resonance $\alphas$
{[}see Eq.~(\ref{eq:pwaveAmplitude_Expansion}){]} may also be called
shortly the \emph{\alphasname}; 
\item[$iii)$] for a short range potential of radius $R$ (in this paper, $R$ is
of the order of $1/\Lambda_{1}$), the \alphasname~verifies the
`width-radius inequality'$\alphas R\gtrsim1$ (the inequality is not
strict, depending on the precise definition chosen for the radius
$R$), which corresponds to the Wigner bound~\cite{Wigner1955,Nishida2012}
imposed by the positivity of the probability density. For the model
potentials used in this study, $\alphas$ is of the order of $\Lambda_{1}$.
In more general situations, for instance for a multichannel interaction,
this parameter can be much larger than $\Lambda_{1}$, corresponding
to a narrow resonance limit; 
\item[$iv)$] In the limit ($1/v\to0^{+}$), there is no p-wave scattering resonance
and there is a shallow p-wave dimer of binding wave number $1/\sqrt{\alphas v}$. 
\end{enumerate}
Hence, the resonant regime in the p-wave scattering differs from the
one in the s wave where the unitary limit can be reached in a large
range of the momentum and not only for a specific value. In what follows,
due to the continuity found in the trimer spectrum at $1/v=0$ (thus
including arbitrarily large and negative or positive scattering volumes),
we will formally qualify this limit as the \emph{p-wave resonance
limit}.

We now turn to the simplification of the STM equation for an isotropic
p-wave interaction. For convenience, we introduce the spinor 
\begin{equation}
\bm{P}(\bm{k})=\sum_{m=-1}^{1}P_{m}(\bm{k})\hat{\bm{e}}_{m}.
\end{equation}
The equations (\ref{eq:STM1}-\ref{eq:STM2}) thus become equations
on ${S(\bm{k})}$ and ${\bm{P}(\bm{k})}$:
\begin{align}
-\frac{\vert\chi(i\kappa_{0})\vert^{2}}{4\pi f_{0,0}(i\kappa_{0})}S(\bm{k})=\qquad\qquad\qquad\qquad\qquad\qquad\label{eq:STM1vec}\\
y\int\frac{d^{3}k^{\prime}}{(2\pi)^{3}}\frac{\langle\chi\vert\bm{k}^{\prime}+\frac{x}{2y}\bm{k}\rangle\langle\bm{k}+\frac{x}{2y}\bm{k}^{\prime}\vert\chi\rangle S(\bm{k}^{\prime})}{y(k^{2}+k^{\prime2})+x\bm{k}\cdot\bm{k}^{\prime}+q^{2}}\nonumber \\
+3\int\frac{d^{3}k^{\prime}}{(2\pi)^{3}}\frac{\langle\chi\vert\bm{k}^{\prime}+\frac{1}{2y}\bm{k}\rangle\langle\bm{k}+\frac{1}{2}\bm{k}^{\prime}\vert\phi\rangle\bm{k}\cdot\bm{P}(\bm{k}^{\prime})}{k^{2}+yk^{\prime2}+\bm{k}\cdot\bm{k}^{\prime}+q^{2}}\nonumber 
\end{align}
\begin{align}
 & \frac{\vert\kappa_{1}\phi(i\kappa_{1})\vert^{2}}{4\pi f_{1}(i\kappa_{1})}\bm{P}(\bm{k})=\label{eq:STM2vec}\\
 & 2y\int\frac{d^{3}k^{\prime}}{(2\pi)^{3}}\frac{\bm{k}\langle\phi(k)\vert\bm{k}^{\prime}+\frac{\bm{k}}{2}\rangle\langle\bm{k}+\frac{\bm{k}^{\prime}}{2y}\vert\chi\rangle S(\bm{k}^{\prime})}{yk^{2}+k^{\prime2}+\bm{k}\cdot\bm{k}^{\prime}+q^{2}}\nonumber 
\end{align}
where we used $\Phi_{m}(\bm{k})=\phi(k)\bm{k}\cdot\hat{\bm{e}}_{m}$.

In this work, we consider the states of total angular momentum $J=1$
and negative parity, consistent with the known symmetry of the KM
trimers. Therefore, when the relative angular momentum between one
fermion and the third particle is zero, the remaining angular momentum
between the fermion-particle pair and the other fermion must be unity.
Assuming that the total momentum projection is zero along a fixed
unit vector $\hat{\bm{e}}_{z}$, $S(\bm{k})$ must be of the form:
\begin{equation}
S(\bm{k})=s(k)\cos\theta\label{eq:Sform}
\end{equation}
where $\theta$ is the angle between $\bm{k}$ and the fixed vector
$\hat{\bm{e}}_{z}$.

Likewise, when the relative angular momentum between the two fermions
is unity, the remaining angular momentum between the two-fermion pair
and the third particle can be either zero or two. It follows that
$\bm{P}(\bm{k})$ is of the form: 
\begin{equation}
\bm{P}(\bm{k})=p_{0}(k)\hat{\bm{e}}_{z}+p_{2}(k)\left[\hat{\bm{e}}_{k}\times(\hat{\bm{e}}_{k}\times\hat{\bm{e}}_{z})\right]\label{eq:Pform}
\end{equation}
where $\hat{\bm{e}}_{k}=\bm{k}/k$ - see Appendix~3 for details.
Inserting Eqs.~(\ref{eq:Sform}) and (\ref{eq:Pform}) into Eqs.~(\ref{eq:STM1vec}-\ref{eq:STM2vec})
yields a set of three integral equations for $s$, $p_{0}$ and $p_{2}$:

\begin{align}
\frac{\vert\chi(i\kappa_{0})\vert^{2}}{f_{0}(i\kappa_{0})} & s(k)+\int_{0}^{\infty}\frac{dk^{\prime}}{\pi}k^{\prime\,2}\Big[L(k,k^{\prime})s(k^{\prime})\nonumber \\
 & +\left(3L_{0}(k,k^{\prime})-L_{2}(k,k^{\prime})\right)p_{0}(k^{\prime})\nonumber \\
 & +\left(L_{2}(k,k^{\prime})-2L_{0}(k,k^{\prime})\right)p_{2}(k^{\prime})\Big]=0\label{eq:STM1-isotropic}
\end{align}
\begin{equation}
3\frac{\vert\kappa_{1}\phi(i\kappa_{1})\vert^{2}}{2yf_{1}(\kappa_{1})}p_{0}(k)-\int_{0}^{\infty}\frac{dk^{\prime}}{\pi}k^{\prime\,2}L_{0}(k^{\prime},k)^{*}s(k^{\prime})=0\label{eq:STM2-isotropic}
\end{equation}
\begin{equation}
3\frac{\vert\kappa_{1}\phi(i\kappa_{1})\vert^{2}}{yf_{1}(\kappa_{1})}p_{2}(k)-\int_{0}^{\infty}\frac{dk^{\prime}}{\pi}k^{\prime\,2}L_{2}(k^{\prime},k)^{*}s(k^{\prime})=0\label{eq:STM3-isotropic}
\end{equation}
The kernels $L$, $L_{0}$, and $L_{2}$ are given by: 
\begin{align}
L(k,k^{\prime}) & =\int_{-1}^{1}du\;yu\frac{\chi^{*}(\vert\bm{k}^{\prime}+\frac{x}{2y}\bm{k}\vert)\chi(\vert\bm{k}+\frac{x}{2y}\bm{k}^{\prime}\vert)}{y(k^{2}+k^{\prime2})+xkk^{\prime}u+q^{2}}\label{eq:L}\\
L_{0}(k,k^{\prime}) & =3\int_{-1}^{1}du\left(ku^{2}+\frac{1}{2}k^{\prime}u\right)\label{eq:L0}\\
 & \qquad\qquad\times\frac{\chi^{*}(\vert\bm{k}^{\prime}+\frac{1}{2y}\bm{k}\vert)\phi(\vert\bm{k}+\frac{1}{2}\bm{k}^{\prime}\vert)}{k^{2}+yk^{\prime2}+kk^{\prime}u+q^{2}}\nonumber \\
L_{2}(k,k^{\prime}) & =3\int_{-1}^{1}du\left(k\left(3u^{2}-1\right)+k^{\prime}u\right)\label{eq:L2}\\
 & \qquad\qquad\times\frac{\chi^{*}(\vert\bm{k}^{\prime}+\frac{1}{2y}\bm{k}\vert)\phi(\vert\bm{k}+\frac{1}{2}\bm{\bm{k}^{\prime}}\vert)}{k^{2}+yk^{\prime2}+kk^{\prime}u+q^{2}}\nonumber 
\end{align}

where $u=\bm{k}\cdot\bm{k}^{\prime}/(kk^{\prime})$ is the cosine
of the angle between $\bm{k}$ and $\bm{k}^{\prime}$.

\subsection{Form factors}

Although the values of the interaction strengths $\xi$ and $g$ can
be set to reproduce a given scattering length and scattering volume
through Eqs.~(\ref{eq:inversea}-\ref{eq:inversev}), the form factors
$\chi$ and $\phi$ remain to be chosen. We consider four different
types of form factors, referred to as the ``Gaussian\textquotedbl ,
``Yamaguchi\textquotedbl , ``Yamaguchi-squared\textquotedbl , and
``Cut-off\textquotedbl{} models. Their expressions are given in Tables~\ref{tab:S-wave-Potentials}
and \ref{tab:P-wave-Potentials} of Appendix~1. The tables also provide
the explicit forms of the two-body scattering amplitudes $f_{0}$
and $f_{1}$, as well as the corresponding parameters $\rs$ and $\alphas$.
In addition to these simple model potentials, a more realistic separable
potential is also considered, which is constructed to reproduce exactly
the wave function scattered at zero energy by a Lennard-Jones potential.
The details of this separable potential have been given in~\cite{Naidon2014a}.

\section{Absence of interaction between the fermions\label{sec:Absence}}

\begin{figure}
\includegraphics[width=8cm]{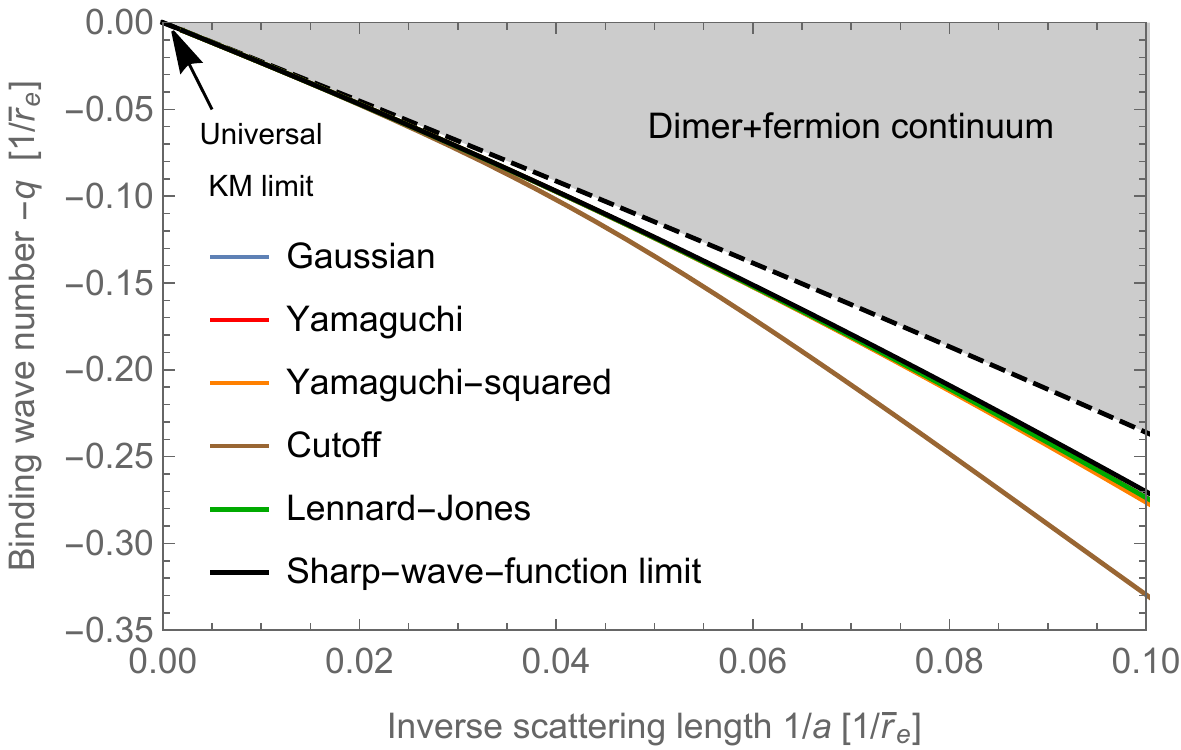}

\includegraphics[width=8cm]{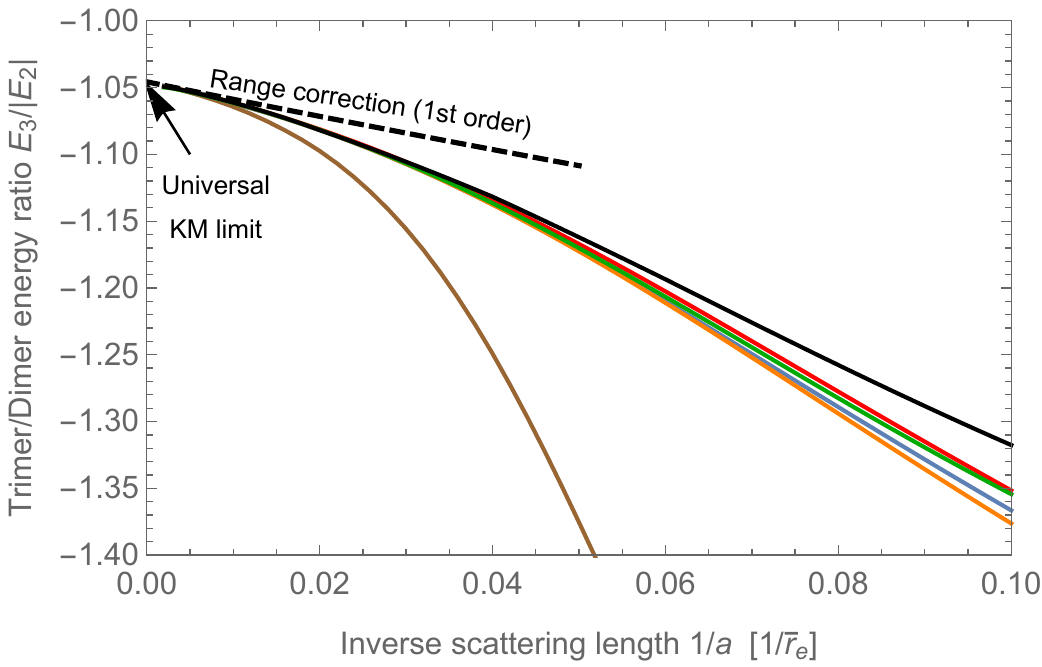}

\caption{\label{fig:noInteraction} Trimer spectrum without fermion-fermion
interaction for a mass ratio ${x=9}$. Top panel: trimer's binding
wave number $q$ as a function of the inverse of the scattering length
$a$ for different models (note that the curves corresponding to some
models cannot be distinguished in this panel - see lower panel). Dashed
curve: threshold of the dimer+fermion scattering continuum (shaded
area) obtained from the binding wave number of the fermion-particle
s-wave dimer in the effective-range approximation (see main text);
Solid curves: trimer binding wave number for different models of the
fermion-particle interaction (see Table~\ref{tab:S-wave-Potentials}
and main text). Bottom panel: ratio between the trimer and dimer energies
$E_{3}/\vert E_{2}\vert$ as a function of the inverse scattering
length; Dashed curve: first-order correction in $\rs/a$ with respect
to the zero-range potential limit (see Eqs.~(\ref{eq:PerturbativeResultDimer}-\ref{eq:PerturbativeResultTrimer})
and main text).}
\end{figure}

\begin{figure}
\includegraphics[width=8cm]{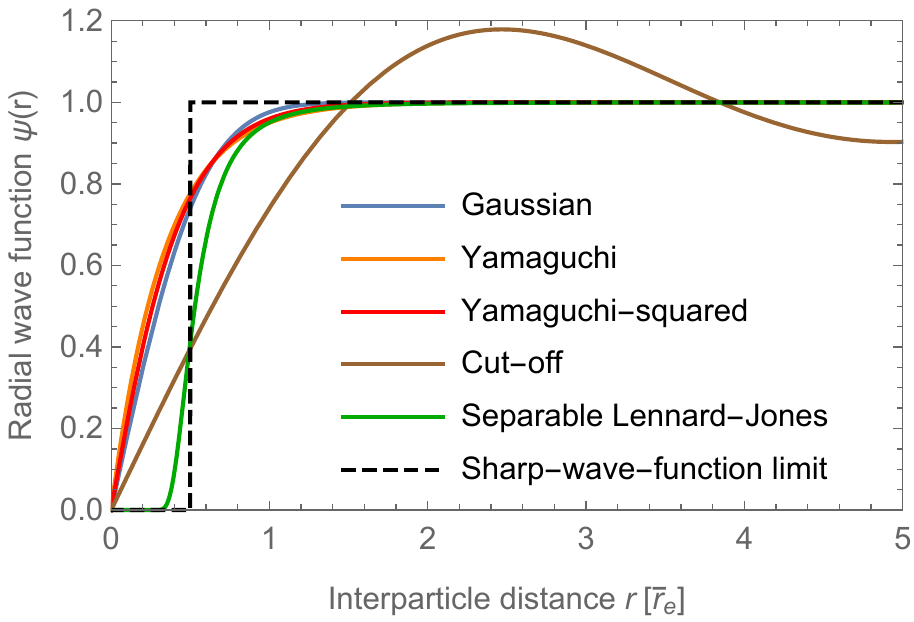}\caption{\label{fig:S-wave-radial-wave}S-wave radial two-body wave function
at zero energy obtained for different models at the s-wave resonance
($|a|\to\infty$) - see Table~\ref{tab:S-wave-Potentials} and main
text. The wave function is normalised to tend to unity asymptotically,
and the distance is expressed in units of the \rsname\, $\rs$.}
\end{figure}

In this section, we consider the regime where the interaction between
the fermions can be neglected. Thus we will use Eq.~(\ref{eq:STM1})
with $P_{m}=0$, that is to say, Eq.~(\ref{eq:STM1-isotropic}) with
$p_{0}=p_{2}=0$.

In this regime, the results are expected to conform to the Kartavtsev-Malykh
universal theory for large enough scattering lengths $a$. Indeed,
in the small momentum and energy limits where $k,k^{\prime},q,1/a\ll\Lambda_{0}$,
Eq.~(\ref{eq:STM1-isotropic}) is equivalent to the zero-range limit
of the STM equation with $\Lambda_{0}\to\infty$ for a fixed value
of $a$. This can be understood as follows: first, consider the small-energy
$q\ll\Lambda_{0}$ and small-momentum limit $k\ll\Lambda_{0}$ of
Eq.~(\ref{eq:STM1-isotropic}). For such small wave numbers, $\vert\chi\vert^{2}/f_{0}$
may be approximated by $-1/a+\kappa_{0}$ according to Eq.~(\ref{eq:swaveAmplitude_Expansion}).
Although the integral over $k^{\prime}$ extends to infinity, i.e.
values of $k^{\prime}$ where $s(k^{\prime})$ is not described by
the small-energy and small-momentum approximation, it turns out that
for $1/a\ll\Lambda_{0}$, $s(k)$ is peaked around $k\sim1/a$ and
then decays to zero. The contributions from $k^{\prime}\gg1/a$ are
thus suppressed by $s(k^{\prime})$ and since $1/a\ll\Lambda_{0}$,
the form factors $\chi$ inside $L(k,k^{\prime})$ may be approximated
to unity. It follows that the small-momentum approximation of the
STM equation is self-consistent and properly describes the limit of
the STM equation for small energy and large scattering length.

This approximated equation formally corresponds to the zero-range
STM equation obtained for two-body contact interactions and no three-body
contact interaction. It is thus equivalent to the Kartavtsev--Malykh
contact theory with a zero three-body parameter. From this, we conclude
that the finite range of the s-wave interactions does not lead to
a non-zero three-body parameter, in contrast to what happens for particles
undergoing the Efimov effect, for which the three-body parameter is
largely set by the finite range of interactions~\cite{Wang2012,Naidon2014,Naidon2016}.
A similar situation was numerically observed before for a zero-range
model with a momentum cutoff equivalent to a repulsive three-body
force~\cite{Endo2012}, while a boundary condition equivalent to
an attractive three-body force seems to set a non-zero three-body
parameter~\cite{Safavi-Naini2013}. We therefore surmise that an
attractive three-body force is necessary for the three-body parameter
defined by Kartavtsev and Malykh to be non-zero.

The first correction with respect to the zero-range limit (relevant
for increasing values of $1/a$) is given by considering the next
order in the small-momentum expansion of $\vert\chi\vert^{2}/f_{0}$.
This leads to the STM equation in the effective-range approximation,
\begin{multline}
\left(-\frac{1}{a}+\kappa_{0}-\frac{\rs}{2}\kappa_{0}^{2}\right)s(k)+\int_{0}^{\infty}\frac{dk^{\prime}}{\pi}\frac{yk^{\prime}}{xk}s(k^{\prime})\\
\biggl[2-\frac{y(k^{2}+k^{\prime2})+q^{2}}{xkk^{\prime}}\\
\times\log\left(\frac{y(k^{2}+k^{\prime2})+q^{2}+xkk^{\prime}}{y(k^{2}+k^{\prime2})+q^{2}-xkk^{\prime}}\right)\biggr]=0.\label{eq:lowMomentumSTM}
\end{multline}
One can treat the range correction $\frac{\rs}{2}\kappa_{0}^{2}$
in this last equation as a perturbation with respect to the STM equation
in the zero-range limit (see Appendix~4). This yields the corrected
energies for the fermion-particle s-wave dimer and the trimer at the
first order in the small parameter $\frac{\rs}{a}$: 
\begin{align}
E_{2} & =E_{2}^{(0)}\left(1+\frac{\rs}{a}\right)\label{eq:PerturbativeResultDimer}\\
E_{3} & =E_{3}^{(0)}\left(1+\frac{\rs}{a}\langle(\kappa_{0}a)^{2}\rangle\right)\label{eq:PerturbativeResultTrimer}
\end{align}
where $E_{2}^{(0)}=y\hbar^{2}/Ma^{2}$ and $E_{3}^{(0)}\propto y\hbar^{2}/Ma^{2}$
are the energies obtained in the zero-range limit, and $\langle\cdots\rangle$
denotes the average in the state $\vert S\rangle$, where $\langle\bm{k}\vert S\rangle=S(\bm{k})=s(k)\cos(\theta)$
corresponds to the eigenvector $s(k)$ of the uncorrected STM equation,
i.e. Eq.~(\ref{eq:lowMomentumSTM}) in the zero-range limit $\rs\to0$.

Figure~\ref{fig:noInteraction} shows the spectrum obtained for the
various model potentials of Table~\ref{tab:S-wave-Potentials}, in
the case of a mass ratio $x=9$, for which only one shallow s-wave
induced trimer state exists for positive scattering length $a$ and
vanishes in the three-body threshold at ${1/a=0}$. In the range of
scattering lengths of the plots, the fermion-particle s-wave dimer
binding wave number ${q_{2}}$ is nearly identical for all models
and is given from Eq.~\eqref{eq:PerturbativeResultDimer} with $q_{2}=\sqrt{-2\mu_{23}E_{2}}/\hbar$.
The dashed curve in the top panel represents $-q_{2}/\sqrt{y}$, which
corresponds to the threshold of the dimer+fermion scattering continuum
(shaded area).

One can distinguish essentially three regions in these plots. The
first region where $\rs/a\ll0.01$ corresponds to the vicinity of
the unitary limit where the zero-range approach used by Kartavtsev
and Malykh applies. In this region, the energy ratio $E_{3}/E_{2}$
between the trimer and dimer is the same for all models and is given
within a few tenths of percent by the zero-range limit $1.0457...$
for the considered mass ratio $x=9$. We note that this universal
region is very narrow and would require a fine tuning of the scattering
length to be observable in ultracold-atom experiments. As expected,
the obtained energy ratio is consistent with the one predicted by
the Kartavtsev-Malykh theory with a three-body parameter equal to
zero.

The second region, where $\rs/a\lesssim0.01$ is universally described
by the scattering length $a$ and the \rsname $\rs$ in agreement
with Eqs.~(\ref{eq:PerturbativeResultDimer}, \ref{eq:PerturbativeResultTrimer}).
For the considered mass ratio, we find $\langle(\kappa_{0}a)^{2}\rangle=2.26\dots$.
This perturbative result, shown as a dashed line in the bottom panel
of Fig.~(\ref{fig:noInteraction}), agrees with all models in this
region within 0.6\%.

In the third region, corresponding to larger values of $\rs/a$, the
trimer energy becomes non-universal. However, we note that for $\rs/a\lesssim0.1$
it remains nearly the same for all models except the cut-off model.
This may be understood by the fact that all these models suppress
the wave function within the range $\rs$ while hardly affecting it
beyond $\rs$. This can be seen in Fig.~\ref{fig:S-wave-radial-wave}.
The zero-energy two-body radial wave function $\psi(r)$ for these
models roughly approaches what we call the sharp-wave-function limit
(dashed curve in Fig.~\ref{fig:S-wave-radial-wave}) given by: 
\begin{equation}
\psi(r)=\begin{cases}
0\qquad\text{(fully suppressed)} & r\le r_{{\rm c}}\\
1-\frac{r}{a}\qquad\text{(free)} & r>r_{{\rm c}}
\end{cases}\label{eq:Sharp-wave-limit}
\end{equation}
This wave function has an effective range that reaches the Wigner
bound~\cite{Wigner1955,Phillips1997,Hammer2010a}, $2r_{{\rm c}}\left(1-\frac{r_{{\rm c}}}{a}+\frac{r_{{\rm c}}^{2}}{3a^{2}}\right)$.
It is generically approached by the zero-energy wave function of deep
short-range potentials supporting many bound states~\cite{Naidon2014}.
It can be exactly obtained from an infinitely deep and narrow potential
well located at some distance $r_{{\rm c}}$, or from a separable
potential with the following form factor: 
\begin{equation}
\chi(k)=\left(1-\frac{r_{{\rm c}}}{a}\right)\cos(kr_{{\rm c}})+\frac{\sin(kr_{{\rm c}})}{ka}\label{eq:Sharp-wave-form-factor}
\end{equation}
The results from this separable potential are shown by the solid black
curves in Fig.~\ref{fig:noInteraction} and are in fair agreement
with other models. The cut-off model stands out as an exception, and
this can be understood from the fact that its radial wave function
is markedly different from the sharp-wave-function limit~(\ref{eq:Sharp-wave-limit}):
it features oscillations at large distances (see Fig.~\ref{fig:S-wave-radial-wave}),
which makes its effective range small compared to its true range $1/\Lambda_{0}$.
Interactions with negative effective range, not considered in this
study, would even more markedly differ from the sharp-wave-function
picture.


\section{\label{sec:Trimers-induced-by}Trimers induced by the p-wave interaction
between the fermions}

In the following part, we consider the model case where the p-wave
interaction is isotropic. We can thus use the symmetry considerations
of section \ref{sec:symmetry}. Since most of the relevant s-wave
models are equivalent for $\rs/a\lesssim0.1$, in this section, the
s-wave interaction between the fermions and the third particle is
described by the Gaussian model.

\subsection{Doubly resonant limit\label{subsec:Doubly-resonant-limit}}

\label{section:doublyResonantLimit}

We first consider simultaneously the s- and p-wave resonant regimes
(${1/a=0}$, ${1/v=0}$) and in order to avoid the Efimov effect,
the spectrum is computed for mass ratios below the critical value
${x<x_{{\rm c}}}$. The limit ${1/a=0}$ ensures that there is no
shallow fermion-particle s-wave dimer nor s-wave induced trimer, and
the limit ${1/v=0}$ ensures that there is no shallow fermion-fermion
p-wave dimer. One can expect that this regime corresponding to $s$-wave
resonant scattering and in the vicinity of the p-wave resonance is
favourable for the occurence of shallow trimers of another type than
the KM or Efimov states.

\begin{figure}
\includegraphics[width=8cm]{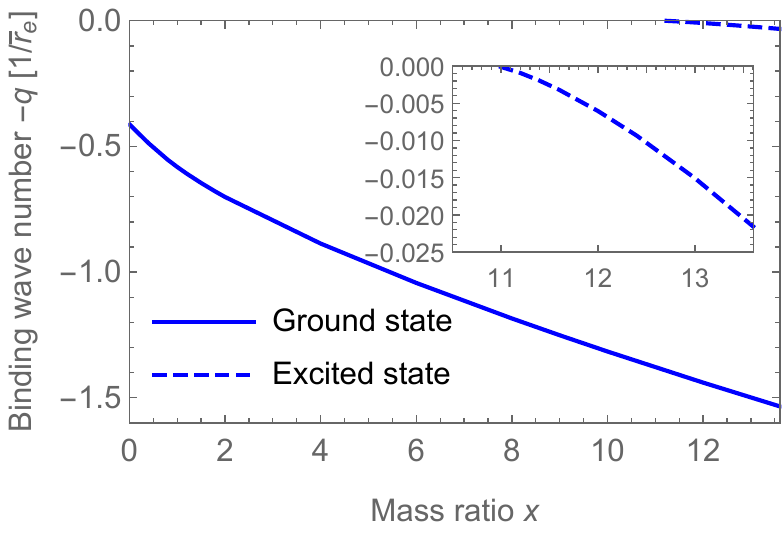} \caption{\label{fig:trimer_s_p_resonance} Trimer spectrum in the presence
of a p-wave interaction between the fermions, at the doubly resonant
limit ${1/a\to0}$ and ${1/v\to0}$, and as a function of the mass
ratio $x$ between the fermions and the third particle. Here, the
p-wave interaction is set to the Gaussian model of Table~\ref{tab:P-wave-Potentials}
of Appendix 1, with ${1/\alphas=\rs}$.}
\end{figure}

The resulting trimer spectrum is shown in Fig.~\ref{fig:trimer_s_p_resonance}
for the p-wave Gaussian model of Table~\ref{tab:P-wave-Potentials}
of Appendix~1. Here, the value of the p-wave interaction range $1/\Lambda_{1}$
has been set such that $1/\alphas=\rs$. This arbitrary choice is
reasonable for a qualitative understanding of the spectrum, since
both lengths are of the order of the range of atomic interactions.
We find one shallow trimer for all values of the mass ratio and another
one for a mass ratio larger than $x_{c}^{\prime}=10.96\dots$. Clearly,
these trimer states are the consequence of the resonant p-wave interaction
that adds to the already present effective attraction between the
two fermions due to the fermion-particle s-wave interaction. This
is why we adopt the denomination \emph{p-wave induced trimers} for
these states. Although the doubly resonant regime is very specific
and not directly relevant to experimental systems, it demonstrates
the existence of shallow trimers that differ from KM and Efimov states.
In what follows, we will change the parameters of the interactions
to consider more relevant regimes in view of possible experimental
studies of the trimer spectrum.

\subsection{Weak interaction between the fermions\label{sec:Perturbative}}

\begin{figure*}
\includegraphics[width=8cm]{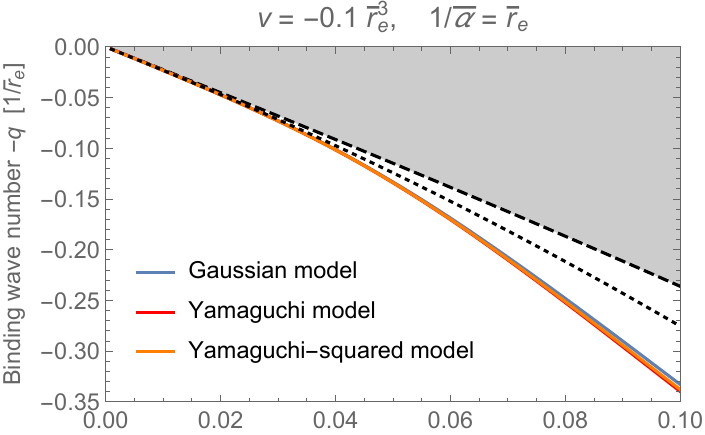} \includegraphics[width=8cm]{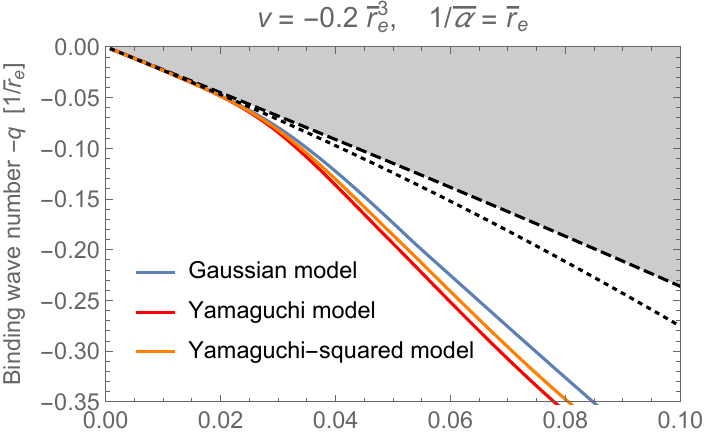}\\
 \caption{\label{fig:weakInteraction}Similar plots to the upper panel of Fig.~\ref{fig:noInteraction}
in presence of a weak attraction between the fermions using the models
of Table~\ref{tab:P-wave-Potentials} of Appendix 1. Left: the scattering
volume is set to $v=-0.1\rs^{3}$; Right: plot for $v=-0.2\rs^{3}$.
Dashed curve: limit of the fermion-particle dimer continuum; Dotted
curve: s-wave induced trimer without fermion-fermion interaction ($v=0$)
for the Gaussian model, corresponding to the blue curve of Fig.~\ref{fig:noInteraction}.}
\end{figure*}

In this section, we aim at understanding how the p-wave induced trimers
emerge from the s-wave induced trimer spectrum of section \ref{sec:Absence}.
We thus consider a gradual increase of the p-wave interaction starting
from the situation where only the fermion-particle s-wave interaction
is present. The mass ratio is fixed at $x=9$ so that without p-wave
interaction there is only one s-wave induced trimer for positive values
of the scattering length.

The p-wave attraction is described by the Gaussian, Yamaguchi, and
Yamaguchi-squared models detailed in Table~(\ref{tab:P-wave-Potentials}).
Similarly to the study of section \ref{sec:Absence}, we have adjusted
the range parameter $\Lambda_{1}$ to reproduce in all models the
same \alphasname $\alphas=1/\rs$. With this choice the diagonal
terms of the STM equation are equal for all models in the small momentum
limit. The spectrum is plotted in Fig.~\ref{fig:weakInteraction}
as a function of $1/a$, for the scattering volumes $v=-0.1\rs^{3},$
and $-0.2\rs^{3}$. For such small scattering volumes, there is no
two-body bound state between the two fermions.

As it can be seen, the trimer's binding is significantly strengthened
even for these weak fermion-fermion attractions. However, as it could
be expected, the KM trimer is not affected by the fermion-fermion
attraction in a region of sufficiently large scattering length $a$.
This region where the universal theory of Kartavtsev and Malykh applies
shrinks with an increasing magnitude of the scattering volume $v$
(compare the left and right panels of Fig.~\ref{fig:weakInteraction}).
One could understand this situation as follows: for a large scattering
length $a$, the KM trimer has a large size (of the order of $a$)
and is mostly not affected by the fermion-fermion interaction, since
it affects the wave function only within a finite range, much smaller
than $a$.

\begin{figure*}
\includegraphics[width=8cm]{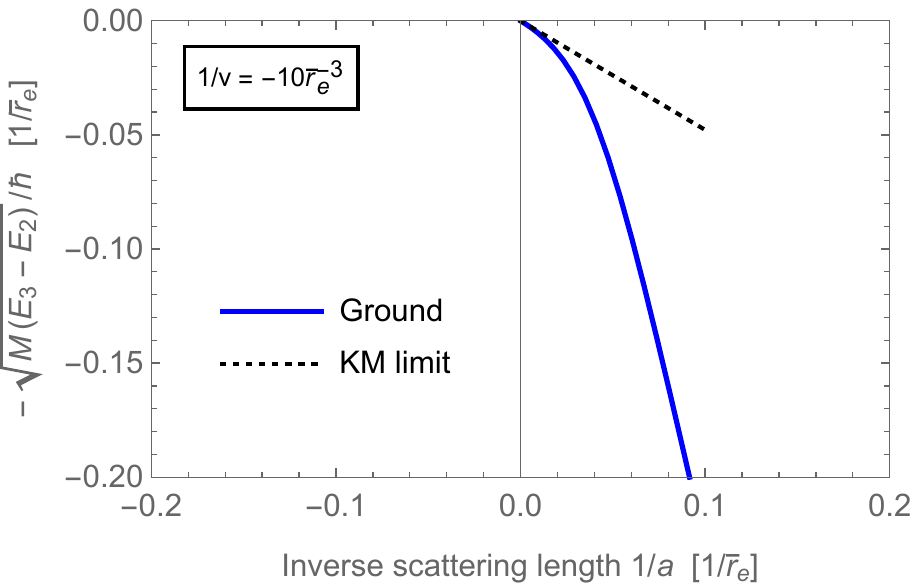}\includegraphics[width=8cm]{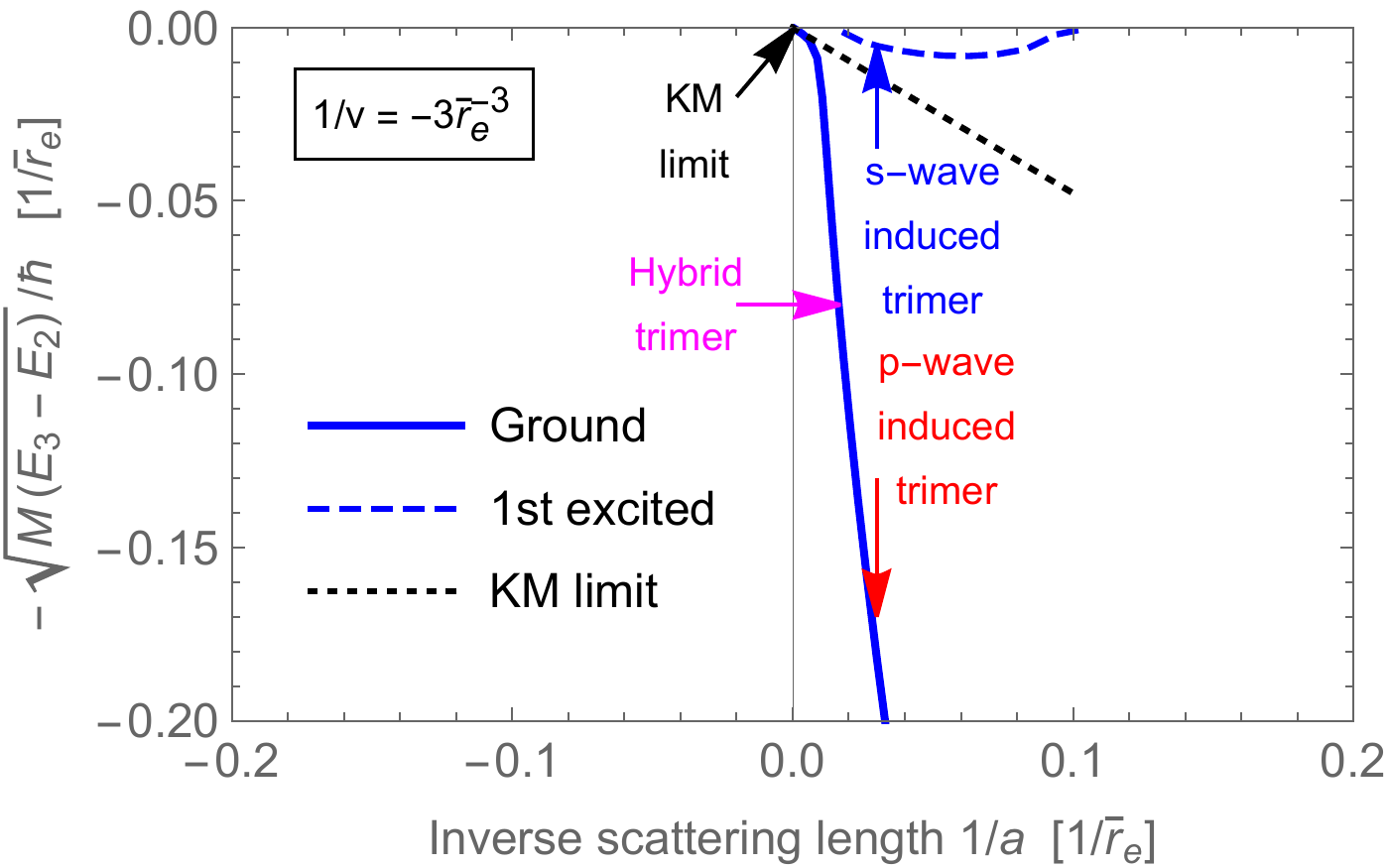}
\includegraphics[width=8cm]{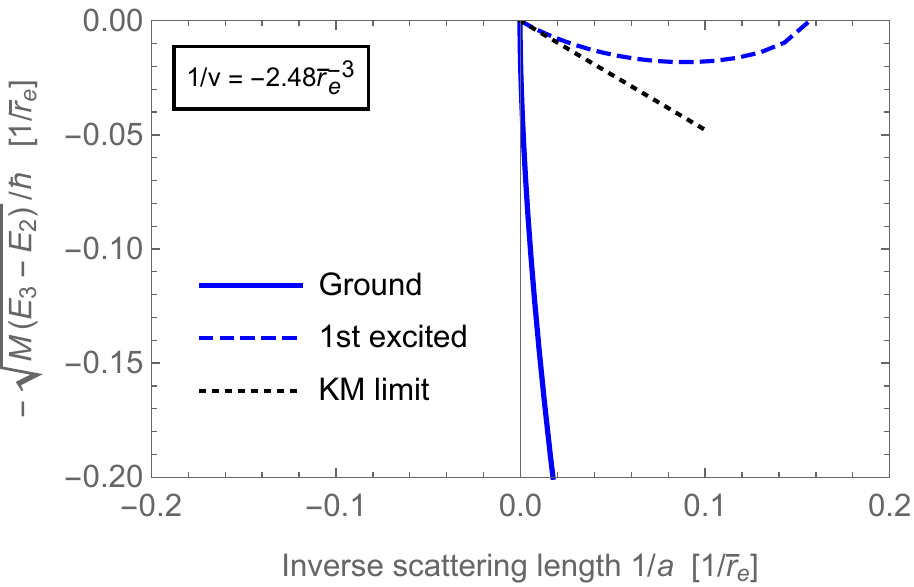}\includegraphics[width=8cm]{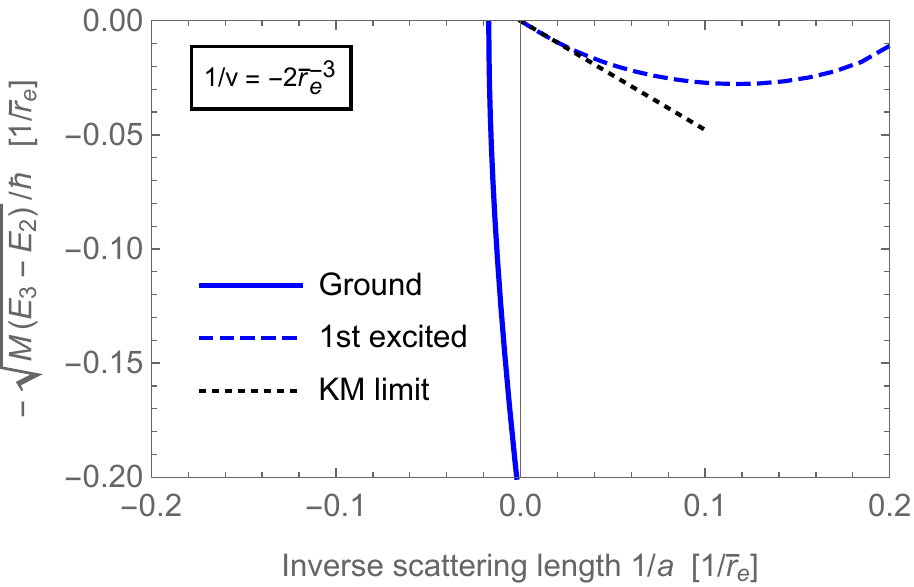}
\includegraphics[width=8cm]{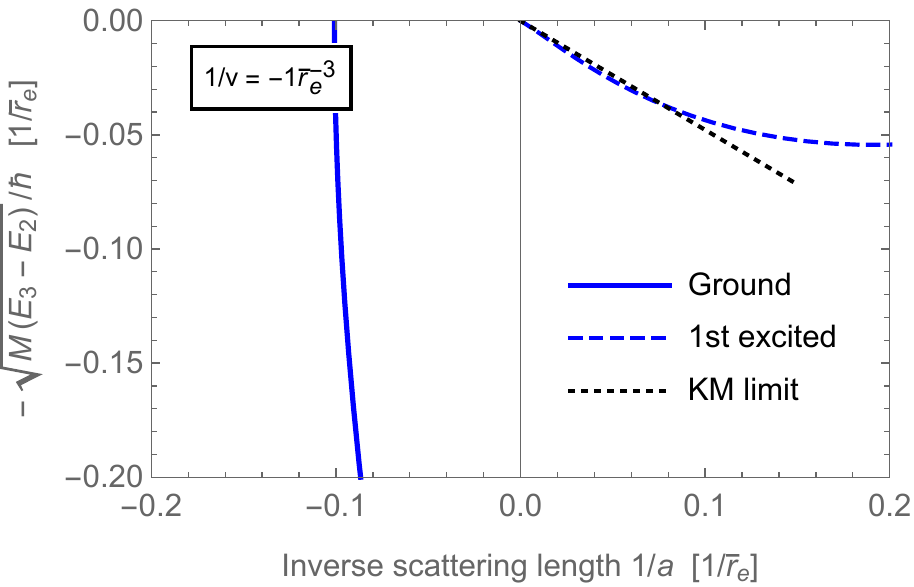}\includegraphics[width=8cm]{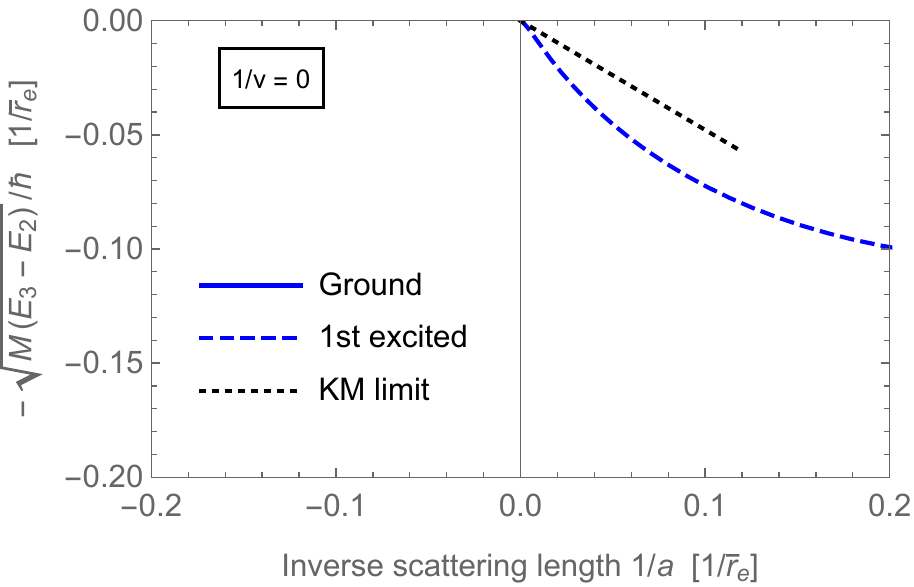}
\caption{\label{fig:Crossing}Dependence of the spectrum on the scattering
volume $v$ between fermions, for increasing values (in magnitude)
of $v$, until a p-wave resonance is reached ($v=-\infty$). The results
are obtained with the p-wave Gaussian model, and both the ground (solid
curve) and excited (dashed curve) trimer states are shown. In order
to distinguish the excited state, the three-body energy is measured
from the two-body energy $E_{2}$ (which is either zero for negative
scattering lengths, or the dimer energy for positive scattering lengths)
and then converted to a wave number. In this fashion, the Kartavtsev--Malykh
zero-range limit corresponds to a straight line shown in black dots.
The arrows indicate the states for which the wave functions of Fig.~\ref{fig:fo_s}
have been computed. The ground state is shown for a wider range of
scattering lengths in Fig.~\ref{fig:ChangingV}.}
\end{figure*}

\begin{figure*}
\includegraphics[width=8cm]{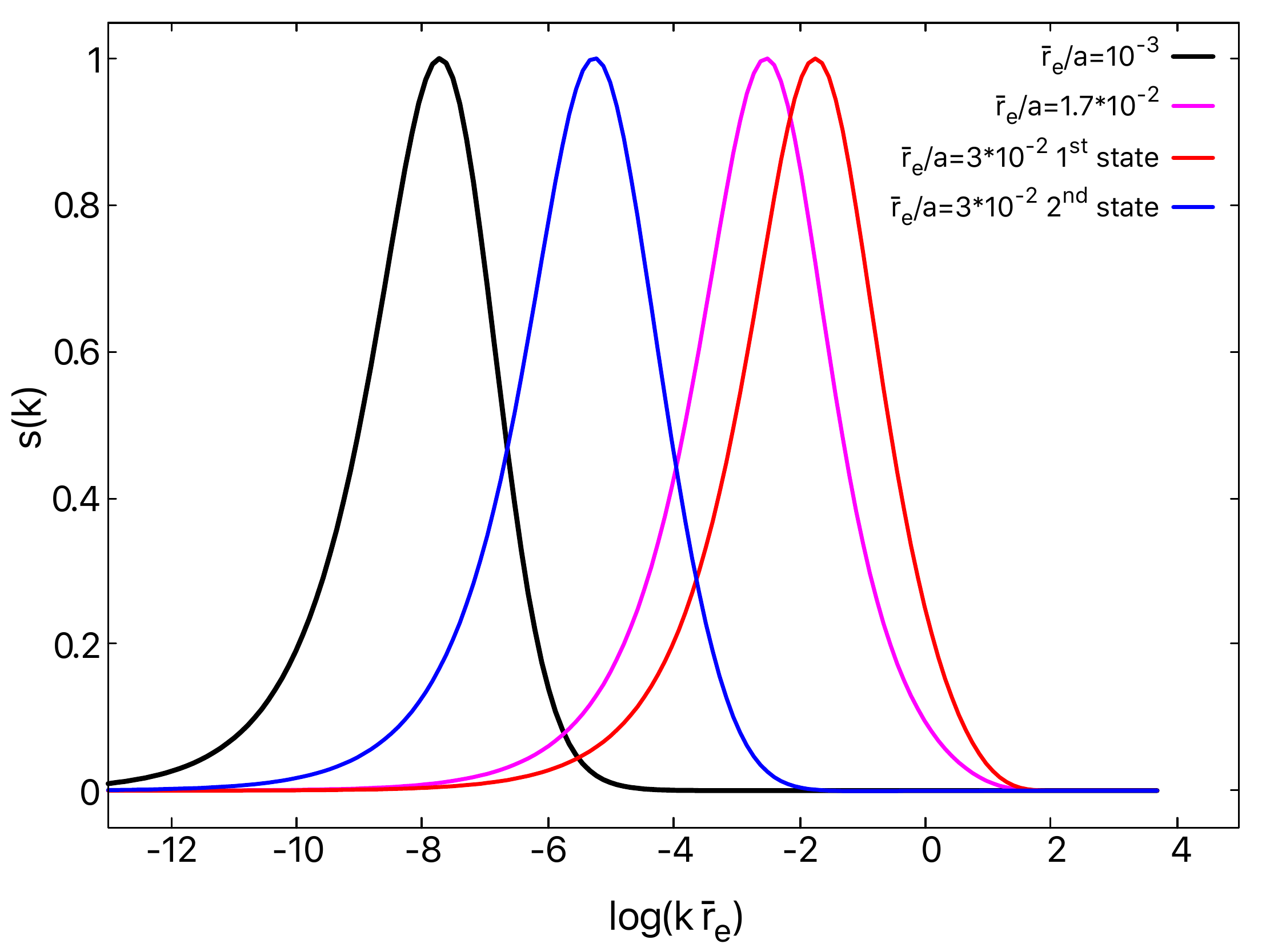} \includegraphics[width=8cm]{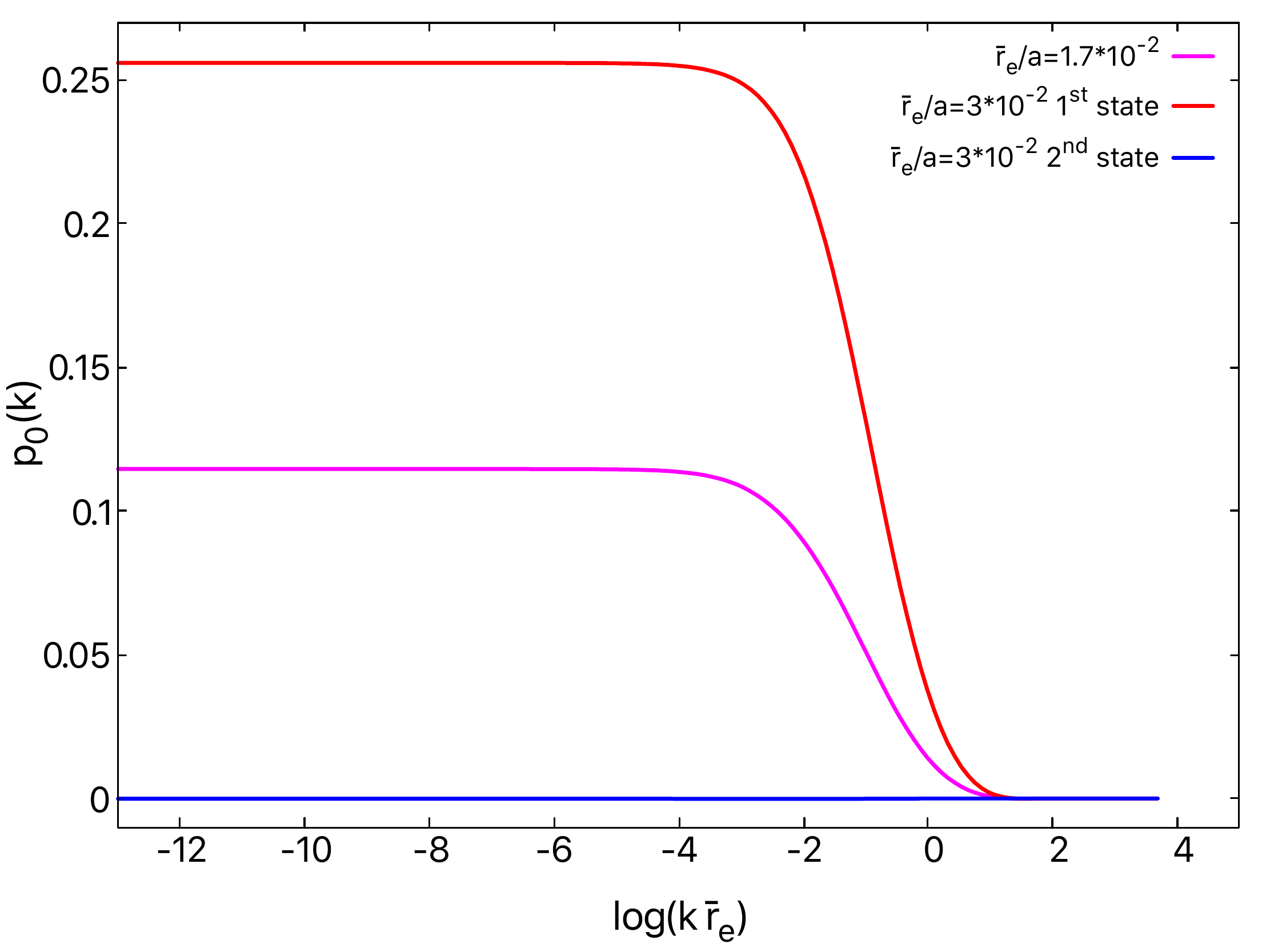}
\caption{\label{fig:fo_s}Components $s$ (left panel) and $p_{0}$ (right
panel) of the wave function in Eq.~(\ref{eq:Pform}) in vicinity
of the crossing between the s-wave and the p-wave induced trimers
for a mass ratio $x=9$ at $\rs^{3}/v=-3$ (see the top right panel
of Fig.~\ref{fig:Crossing}). The component $p_{2}$ is negligible
and not plotted (see section \ref{sec:s-wave-coupling-approximation}).
For $\rs/a=10^{-3}$ (well before the crossing), the s-wave induced
trimer (black) is indistinguishable from the KM state. For $\rs/a=1.7\times10^{-2}$,
the s-wave induced trimer (magenta) hybridises with the (virtual)
p-wave induced trimer. At $\rs/a=3\times10^{-2}$ (after the crossing)
there are two trimers: the first state (ground state) shown in red
and the second state (excited state) shown in blue. }
\end{figure*}

\begin{figure}
\includegraphics[width=8cm]{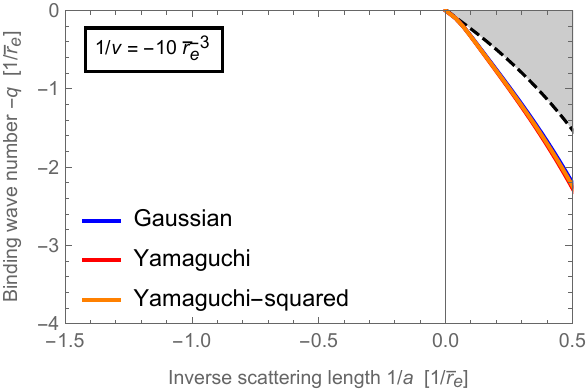} \\
\includegraphics[width=8cm]{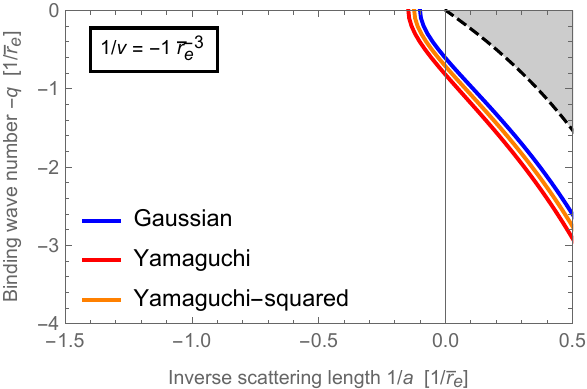} \includegraphics[width=8cm]{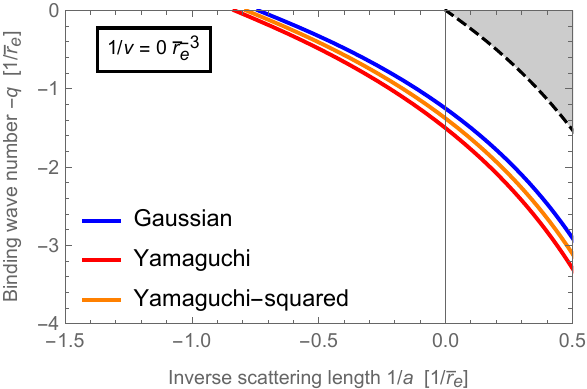}
\caption{\label{fig:ChangingV}Dependence of the spectrum on the scattering
volume $v$ between fermions, for increasing values (in magnitude)
of $v$, until a p-wave resonance is reached ($v=-\infty$). The plots
are similar to the top panel of Fig.~\ref{fig:noInteraction}, although
the different curves now correspond to different models of the p-wave
interaction between the fermions. The value of the \alphasname $\alphas$
is fixed to $1/\rs$ in all models. On the scale of these plots, only
the ground trimer is visible, the excited trimer being indistinguishable
from the dimer (dashed curve).}
\end{figure}

\begin{figure*}
\includegraphics[width=8cm]{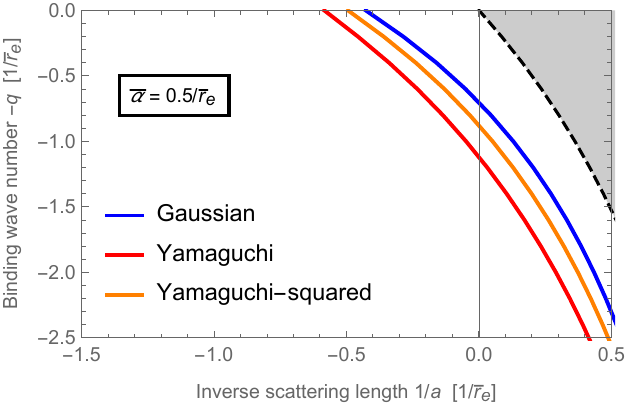} \includegraphics[width=8cm]{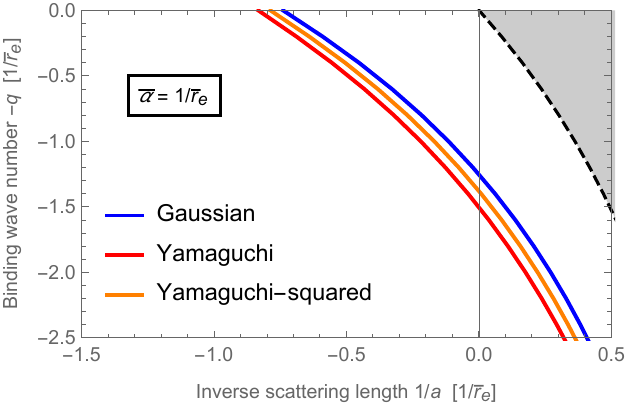}\\
\includegraphics[width=8cm]{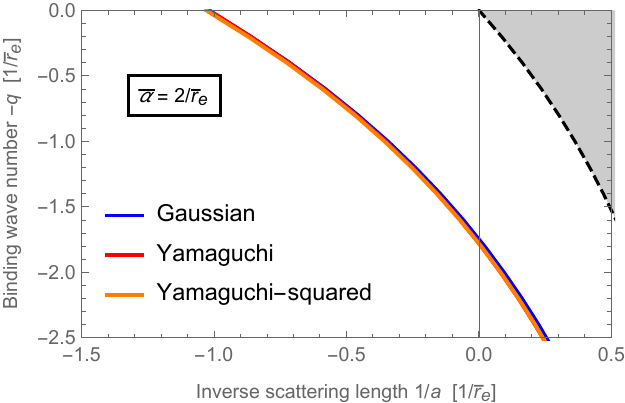} \includegraphics[width=8cm]{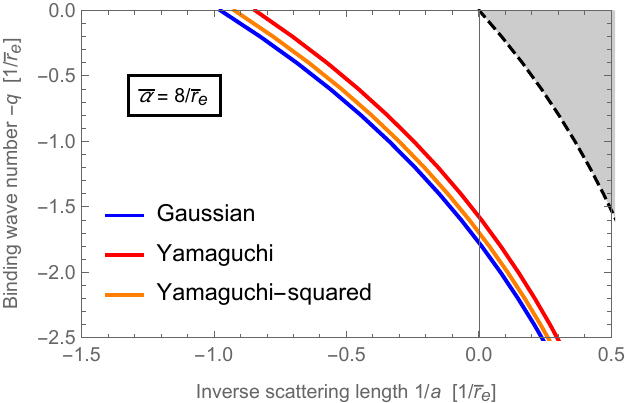}
\caption{\label{fig:PwaveResonance}At the p-wave resonance $v=\infty$. Plots
similar to Fig.~\ref{fig:ChangingV} where, instead of the scattering
volume $v$, the \alphasname $\alphas$ is varied as indicated in
the box of each panel. Note that the fifth panel is the same as the
last panel of Fig.~\ref{fig:ChangingV}. }
\end{figure*}

\subsection{\label{subsec:Critical-scattering-volume}Critical scattering volume
at s-wave resonance}

For increasing strengths of the fermion-fermion interaction, a radical
change occurs in the shallow trimer spectrum as can be seen in Fig.~\ref{fig:Crossing}.
First, as the s-wave induced trimer gets more bound and deviates increasingly
from the universal KM state, a new trimer branch appears from the
s-wave dimer threshold at finite values of the scattering length.
Then, at some critical scattering volume $v_{c}$, the largest scattering
length at which this excited trimer state appears becomes infinite.
Concurrently, the ground trimer energy is pushed down so much that
it conforms to the KM state limit only at $1/a=0$. Past this critical
strength, the ground trimer becomes borromean - it appears from a
negative scattering length - while the excited trimer's threshold
remains fixed at $1/a=0$. As the fermion-fermion attraction is further
increased, it is now the excited trimer which follows the KM state
limit, in an increasingly wide range of scattering lengths, until
this range shrinks again to zero as the scattering volume approaches
infinity. For stronger attraction between the fermions, the scattering
volume becomes positive and there exists a p-wave two-body bound state,
which sets a negative-energy threshold for the occurence of the three-body
bound states. Thus, the KM limit does not exist any more in this regime.

We interpret the interplay between the ground and excited trimers
as a level repulsion and avoided crossing between the s-wave induced
trimer and the p-wave induced trimer states. The analysis of the wave
function at $\rs^{3}/v=-3$ shown in Fig.~\ref{fig:fo_s} confirms
that the s-wave induced trimer hybridises with the p-wave induced
trimer. Before the crossing ($\rs/a=10^{-2}$), $p_{0}$ is negligible
and the wave function $s(k)$ coincides with the solution without
p-wave interaction (cross symbols). After the crossing ($\rs/a=3\times10^{-2}$),
the ground state has a large component $p_{0}(k)$ and can be considered
as the p-wave induced trimer, whereas the excited state has a negligible
component $p_{0}(k)$ and can thus be considered as the s-wave induced
trimer. At the crossing ($\rs/a=1.7\times10^{-2}$) where there is
still one trimer state, the component $p_{0}(k)$ is not negligible
as a consequence of the hybridisation. Thus for increasing values
of $1/a$ the ground branch of the s-wave induced trimer continuously
transforms into the p-wave induced trimer.

As can be seen in Fig.~\ref{fig:ChangingV}, the regime where the
ground-state trimer is borromean is not universal: different models
with the same parameters $v$ and $\alphas$ give different results.
Nevertheless, these results remain qualitatively consistent. Figure~\ref{fig:ChangingV}
shows the dependence on the scattering volume for a fixed effective
range. There is also a dependence on the effective range. This is
shown in Fig.~\ref{fig:PwaveResonance}, at the fixed scattering
volume $v=\infty$ corresponding to the p-wave resonance, i.e. the
occurence of a p-wave two-body bound state between the two fermions.
One can see more marked differences between the models for a p-wave
\alphasname $\alphas$ that is small compared to $1/\rs$, meaning
that we consider a p-wave potential of large range with respect to
the s-wave potential ($\Lambda_{1}\ll\Lambda_{0}$). At the particular
value $1/\alphas=\frac{1}{2}\rs$, all models appear to coincide,
although this is specific to this value and we have no particular
explanation for this coincidence.

Let us now consider the critical scattering volume at which the fermion-fermion
interaction is sufficiently attractive for supporting a borromean
p-wave induced trimer. This occurs at the s-wave resonance ($1/a=0$)
of the fermion-particle interaction, where the s-wave dimer appears.
As shown in Fig.~(\ref{fig:Crossing}), for the mass ratio $x=9$,
using the p-wave Gaussian model, we find $1/v=-2.48/\rs^{3}$ with
the choice $\alphas=1/\rs$.

\begin{figure}
\includegraphics[width=8cm]{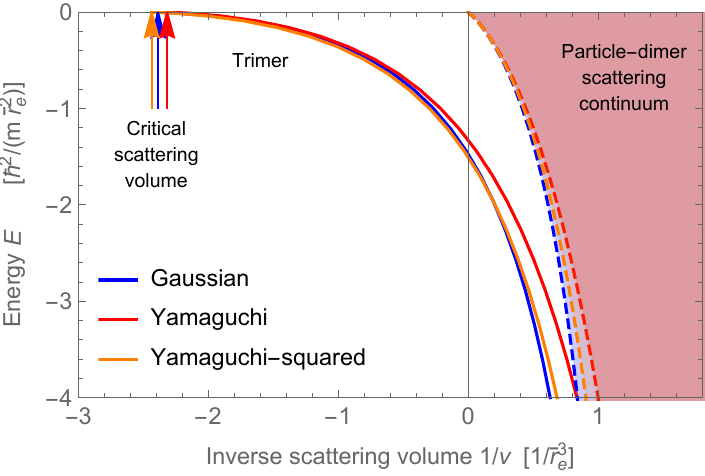}

\caption{\label{fig:ScatteringVolumeDependence} Three-body spectrum as a function
of the inverse scattering volume $1/v$ between the two fermions,
for a fermion-particle interaction at the s-wave unitarity ($a=\infty$)
and mass ratio $x=9$. The solid curves indicate the trimer energy
for different models and the dashed curves indicate the p-wave fermion-fermion
dimer energy. The arrows indicate the critical scattering volume $v_{c}$
at which the trimer appears from the three-body scattering threshold.
This critical scattering volume corresponds to the value beyond which
the trimer is borromean with respect to the fermion-particle dimer,
as seen in Fig.~\ref{fig:ChangingV}.}
\end{figure}

Figure~\ref{fig:ScatteringVolumeDependence} shows the ground-state
trimer energy at the s-wave unitarity as a function of the inverse
scattering volume for three different models satisfying $\alphas=1/\rs$.
The critical scattering volume $v_{c}$ is indicated by an arrow for
each model. As it can be seen, the critical scattering volume is not
universal, although it is close to $1/v_{c}\approx-2.5/\rs^{3}$ for
the three models.

\begin{figure}
\includegraphics[width=8cm]{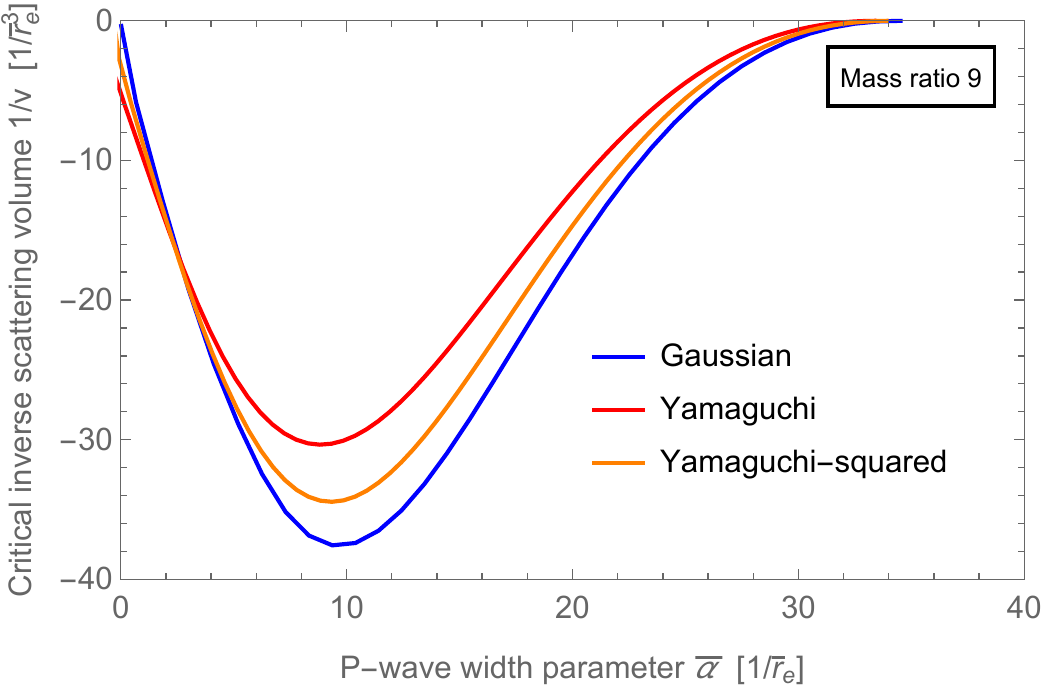}

\includegraphics[width=8cm]{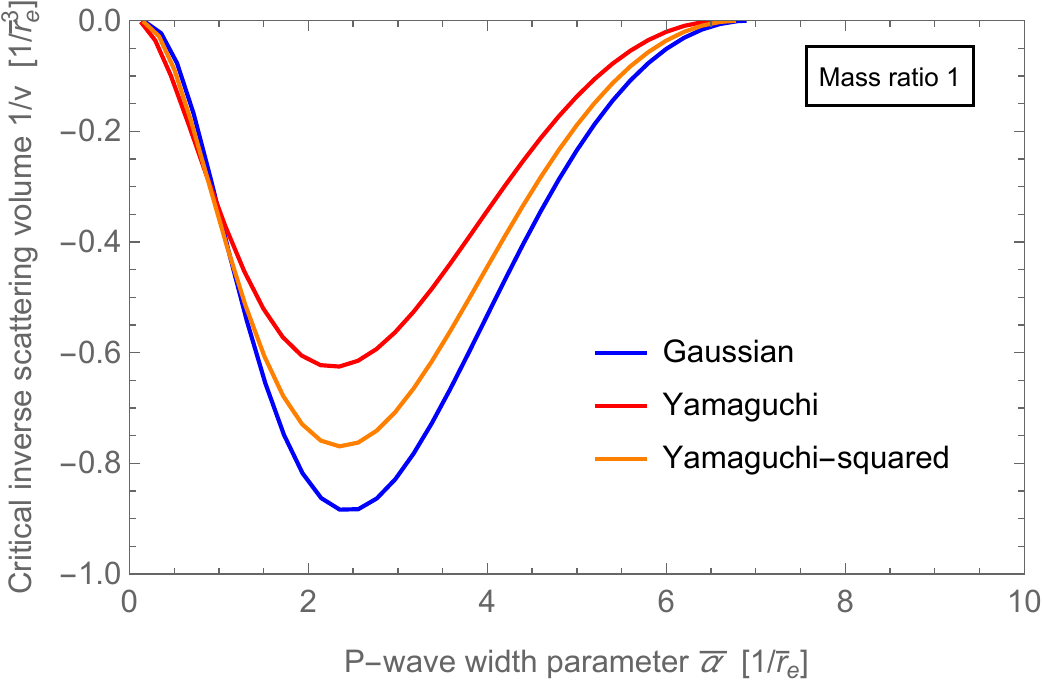}

\caption{\label{fig:CriticalScatteringVolume}Critical inverse scattering volume
$1/v_{c}$, as a function of the \alphasname $\alphas$. Top panel:
for a mass ratio 9. Bottom panel: for a mass ratio 1.}
\end{figure}

The critical scattering volume itself depends on the interaction range
between the fermions, which is of the order of $1/\alphas$. This
dependence is shown in the top panel of Fig.~\ref{fig:CriticalScatteringVolume}
for the mass ratio $x=9$. Again, the results are not model-independent
but they are sufficiently close to draw general conclusions. One can
see that the critical inverse scattering volume vanishes for $1/\alphas\approx\rs/34$
and has a maximum magnitude around $1/\alphas\approx\rs/10$. This
means that the p-wave induced trimer usually exists, unless the range
$1/\alphas$ of the fermion-fermion interaction happens to be more
than 30 times smaller than the range of the particle-fermion interaction.
For a fermion-fermion interaction range $1/\alphas$ of the same order
as the fermion-particle interaction range $\rs$, the critical scattering
volume required to observe the p-wave induced trimer is about $-\rs^{3}/10$,
which means that the fermion-fermion interaction remains modest and
far from the p-wave resonance $v\alphas^{-3}\gg1$.

So far, we have looked at the specific mass ratio $x=9$, since in
the absence of fermion-fermion interaction, KM trimers only exist
for $M/m>x_{1}=8.17260\dots$. However, as we pointed out in Section~\ref{section:doublyResonantLimit},
in the presence of fermion-fermion interaction, the p-wave induced
trimers may exist for any mass ratio. The bottom panel of Fig.~\ref{fig:CriticalScatteringVolume}
shows the critical scattering volume for a mass ratio 1 (all three
particles having the same mass, such as identical atoms in different
spin states). The figure shows that the p-wave induced trimer still
exists, however the required critical scattering volume is larger
than for a mass ratio of 9. For a \alphasname $\alphas$ of the order
of $1/\rs$, the scattering volume must be larger than $2.5\rs^{3}$,
which is closer to the p-wave resonance.

Finally, we should mention that for small positive scattering lengths
or volumes, the models used in this study lead to the presence of
a series of deep trimer states that emerge successively from the dimer
threshold. These deep trimers are due to the usual fact that strong
attractive interactions bind more strongly three particles than two
particles. Unlike the shallow trimers discussed above, these trimers
are very model-dependent. Furthermore, for such deep states the models
do not correpond to any accurate description of a real system. For
these reasons, we do not consider these deep trimers. It is nonetheless
interesting to note that as the system approaches the p-wave resonance,
the excited state shown in Fig.~(\ref{fig:Crossing}) at some point
merges with the highest of these deep states, in a region of small
positive scattering length outside the range of Fig.~(\ref{fig:Crossing}))
. Namely, the scattering length at which the excited state disappears
in the dimer threshold meets the scattering length at which the deep
state appears from the threshold, and the two states become a single
state lying below the dimer threshold.

\section{Non isotropic p-wave interaction}

In the presence of a magnetic field along the $z$ direction, the
p-wave interaction is not isotropic: a phenomenon observed in experiments
\cite{Ticknor2004,Waseem2016}. This results in a difference between
the scattering amplitude $f_{1,0}$ and the scattering amplitudes
$f_{1,1}=f_{1,-1}$ thus leading to a splitting of the isotropic trimer
branches (degeneracy of order three) into two branches. In principle,
this makes the formalism more involved as the total orbital momentum
of the three particles is no more a good quantum number. However the
magnetic quantum number of the total orbital momentum and the parity
are still good quantum numbers of the system. The S function is then
the superposition of odd momentum states with the desired projection
on the ${z}$-axis (${M=0,\pm1}$). The direct diagonalisation of
the STM equation is then a bit tedious.

As we show in what follows, we will use an accurate approximation
of the wave function that permits us to greatly simplify the problem
in a perturbative approach.

\subsection{S-wave coupling approximation\label{sec:s-wave-coupling-approximation}}

For an isotropic p-wave interaction, we have observed that the wave
functions $\bm{P}(\bm{k})$ obtained numerically predominantly consist
of the s-wave state for the relative motion between the impurity and
the pair of identical fermions in a p-wave state, meaning that $\vert p_{2}(k)\vert\ll\vert p_{0}(k)\vert$
and thus 
\begin{equation}
\mathbf{P}(\bm{k})\sim p_{0}(k)\hat{\mathbf{e}}_{z}.\label{eq:s-wave_approx}
\end{equation}
Physically, this can be understood by the fact that the coupling with
the d-wave state is suppressed due to the d-wave centrifugal barrier.
In what follows, Eq.~(\ref{eq:s-wave_approx}) will be called the
\emph{s-wave coupling approximation}.

Inspection of the STM equations (\ref{eq:STM1-isotropic}-\ref{eq:STM3-isotropic})
shows that this corresponds to neglecting the terms involving the
kernel $L_{2}$. One can verify indeed that in the region ${k\rs<1}$,
\begin{equation}
\left|\frac{\int_{0}^{\infty}dk^{\prime}k^{\prime\,2}L_{2}(k,k^{\prime})s(k^{\prime})}{\int_{0}^{\infty}dk^{\prime}k^{\prime\,2}L_{0}(k,k^{\prime})s(k^{\prime})}\right|\ll1.\label{eq:ratioOfIntegrals}
\end{equation}
This last property, which depends in principle on the particular form
of the function $s(k^{\prime})$, is very well satisfied for the solutions
of the STM equations, which decrease quickly for $k\rs\gtrsim1$.
This can be explained qualitatively by the fact that in the domain
$k\rs,k^{\prime}\rs\ll1$, the kernel $L_{2}$ may be approximated
by its zero-range limit where the functions $\chi$ and $\phi$ are
replaced by the unit number: 
\begin{equation}
L_{2}(k,k^{\prime})\sim3\int_{-1}^{1}du\frac{k\left(3u^{2}-1\right)+k^{\prime}u}{k^{2}+yk^{\prime2}+kk^{\prime}u+q^{2}}\label{eq:L2_approx}
\end{equation}
which is exactly zero in the limit ${k^{2}+yk^{\prime2}+q^{2}\gg kk^{\prime}}$.
This limit is indeed always verified in the domain ${k\ll q}$, where
the function ${p_{0}(k)}$ is almost constant. In particular, it is
satisfied for sufficiently deep states ($q\rs$ of the order of the
unity), which is typically the case for the p-wave induced trimers.
For s-wave induced trimers, after a plateau the function $p_{0}$
decreases quickly for $k>q$ (see for example typical shapes of this
function in Fig.~\ref{fig:fo_s}), and the s-wave coupling approximation
is thus relevant to compute these states. For p-wave induced trimers,
we observe that $p_{0}(k)$ is almost constant for $k\lesssim\Lambda_{0}\sim\Lambda_{1}$,
even near the threshold. Outside this plateau, the s-wave coupling
approximation is not satisfied in general but $p_{0}(k)$ becomes
negligible. Hence, we conclude that the s-wave coupling approximation
has a very wide regime of validity.

In the s-wave coupling approximation, the matrix STM equations can
be written as 
\begin{align}
\frac{\vert\chi(i\kappa_{0})\vert^{2}}{f_{0}(i\kappa_{0})} & s(k)+\int_{0}^{\infty}\frac{dk^{\prime}}{\pi}k^{\prime\,2}\Big[L(k,k^{\prime})s(k^{\prime})\nonumber \\
 & +3L_{0}(k,k^{\prime})p_{0}(k^{\prime})\Big]=0,\label{eq:STM1--swaveApprox}
\end{align}
\begin{equation}
3\frac{\vert\kappa_{1}\phi(i\kappa_{1})\vert^{2}}{2yf_{1}(\kappa_{1})}p_{0}(k)-\int_{0}^{\infty}\frac{dk^{\prime}}{\pi}k^{\prime\,2}L_{0}(k^{\prime},k)^{*}s(k^{\prime})=0.\label{eq:STM2--swaveApprox}
\end{equation}
In the next section, we will use the s-wave coupling approximation
to treat the anisotropic p-wave interaction perturbatively.

\subsection{Perturbative approach}

For a given value of the quantum number $M$, we rewrite Eq.~\eqref{eq:STM2}
in the form 
\begin{align}
 & \frac{\left[\kappa_{1}\phi(i\kappa_{1})\right]^{2}\bm{P}({\bm{k}})}{4\pi f_{1,M}(i\kappa_{1})}-2y\int\frac{d^{3}k^{\prime}}{(2\pi)^{3}}\,\biggl[\langle\phi|\bm{k}^{\prime}+\frac{\bm{k}}{2}\rangle\nonumber \\
 & \times\frac{(\bm{k}^{\prime}+\frac{\bm{k}}{2})S({\bm{k}^{\prime}})\langle\frac{\bm{k}^{\prime}}{2y}+\bm{k}|\chi\rangle}{k^{\prime}\,^{2}+yk^{2}+q^{2}+{\bm{k}}\cdot{\bm{k}^{\prime}}}\biggr]+\delta\bm{X}_{M}=0.\label{eq:STM_P_bis}
\end{align}
In Eq.~(\ref{eq:STM_P_bis}), we have introduced the perturbation
${\delta\mathbf{X}_{M}}$ with respect to the formally isotropic STM
equations where we have set ${g\equiv g_{M}}$, ${\phi\equiv\phi_{M}}$
and ${f_{1}(k)\equiv f_{1,M}(k)}$: 
\begin{equation}
\delta\mathbf{X}_{M}=\sum_{m=-1}^{1}\left\{ \frac{\left[\kappa_{1}\phi_{m}(i\kappa_{1})\right]^{2}}{f_{1,m}(i\kappa_{1})}-\frac{\left[\kappa_{1}\phi_{M}(i\kappa_{1})\right]^{2}}{f_{1,M}(i\kappa_{1})}\right\} P_{m}\hat{\mathbf{e}}_{m}.
\end{equation}
At the lowest order, i.e. if one neglects the perturbation, the s-wave
coupling approximation gives the form of the $S$ and $\bm{P}$ functions:
\begin{equation}
S(\bm{k})=\hat{e}_{\bm{k}}\cdot\hat{e}_{M}s(k);\quad P_{m}(\bm{k})=p_{0}(k)\delta_{M,m}.\label{eq:ansatz_anisotropic}
\end{equation}
We can then use the perturbation formalism of Appendix~4. The unperturbed
eigenvector is 
\begin{equation}
\langle\bm{k}|\lambda_{0}\rangle=[s(k)(\hat{e}_{\bm{k}}\cdot\hat{e}_{M}),p_{0}(k)\delta_{M,m}]
\end{equation}
and the shift in energy at the first order of the perturbation is
obtained from the matrix element 
\begin{equation}
\langle\lambda_{0}|\mathcal{M}^{(1)}|\lambda_{0}\rangle=\sum_{m=-1}^{1}\int\frac{d^{3}k}{(2\pi)^{3}}(\hat{e}_{m}^{*}\cdot\delta\bm{X}_{M})P_{m}(\bm{k}).
\end{equation}
Using Eq.~\eqref{eq:ansatz_anisotropic}, one finds that this term
is exactly zero and thus that at the first order of perturbation,
the formally isotropic model gives the branches of the spectrum corresponding
to the magnetic number $m$.

To conclude this section, using the s-wave coupling approximation,
the spectrum indexed by the quantum number $M$ is obtained by considering
for each value of $M$ the system of equations (\ref{eq:STM1--swaveApprox},\ref{eq:STM2--swaveApprox})
where $f_{1}(i\kappa_{1})$ is replaced by $f_{1,M}(i\kappa_{1})$.
The results derived in the isotropic case can thus be directly used
to determine the spectrum for an anisotropic p-wave interaction.

\section{Experimental observation}

The three-body spectrum presented in this study can be probed experimentally
in ultra-cold experiments with standard techniques used for the study
of Efimov trimers. The traditional and most straightforward technique
consists in preparing an ultra-cold mixture of atoms in a certain
state of interaction, and measure the three-body losses by imaging
the cloud of atoms~\cite{Kraemer2006,Ottenstein2008,Zaccanti2009,Pollack2009,Gross2009}.
The variation of the loss rate as a function of a parameter controlling
the interactions should reveal features related to the spectrum. For
instance, near the appearance threshold of the Borromean trimer presented
above, a peak in the losses by three-body recombination is expected.
The detailed shape of this peak has not been addressed here.

Alternatively, depending on the considered atomic species, it is possible
to use the radio-frequency spectroscopy technique, which consists
in preparing the atoms in different spin states than those for which
the interactions support the trimers of interest described in this
study, and induce a spin transition which target those trimers, and
thus measure their energy~\cite{Machtey2012,Lompe2010a,Nakajima2011}.
Such measurements could directly test our theoretical predictions.

However, the first challenge in such experiments is to find and prepare
two atomic species that have a suitable mass ratio and whose two-body
s-wave and p-wave interactions happen to be, or can be controlled
to be, in a region of scattering length and scattering volume where
interesting trimers states are present. This work shows that the original
constraint on the mass ratio for the KM states, which limits the choice
mostly to mixtures of chromium and lithium atoms, can be relaxed provided
there is a sufficiently strong p-wave interaction between the two
fermionic atoms. Finding the most promising candidates requires a
precise knowledge of the interspecies s-wave resonances of mass imbalanced
mixtures, as well as their intraspecies p-wave resonances.

\section{Conclusion\label{sec:Conclusion}}

In this work, we have clarified how the universal trimer bound states
of two identical fermions and one particle discovered by Kartavtsev
and Malykh occur in a realistic setting where finite-range interactions
are present between all particles. We found that in the absence of
a fermion-fermion dimer, the Kartavtsev--Malykh universal trimers
are almost always present for a sufficiently large scattering length
between the fermions and the particle. However, this range of scattering
length is very narrow, making the experimental observation of these
universal states challenging. The trimers extend away from this universal
region to smaller scattering lengths, where they become significantly
more bound. Moreover, we found that the spectrum is enriched by up
to two additional trimers for sufficiently strong attraction between
the fermions. These trimers are borromean, i.e. they can exist even
though the interactions are not sufficiently attractive to bind the
two fermions or bind the particle with one of the two fermions. Although
this extended spectrum is not strictly universal, it follows a generic
scenario as a function of the low-energy parameters of the interactions.
This scenario results from an avoided crossing between trimers induced
only by the s-wave interaction between the fermions and the particle,
and trimers induced by the p-wave interaction between the two fermions.
A striking feature of the borromean trimers is that their ground state
exists for any mass ratio, unlike the universal trimer states whose
existence are limited to mass ratios larger than $x_{1}=8.17260\dots$.
This dramatic enhancement of the range of existence, in terms of both
interaction and mass ratio, makes these trimers much more accessible
to experimental observation.

It is interesting to mention that a shallow borromean trimer in the
same symmetry sector ${J^{\pi}=1^{-}}$ has also been predicted in
the case of three fully polarised fermions experiencing a p-wave resonant
pairwise interaction \cite{Jona-Lasinio2008}. We should also note
that our conclusions are limited to the case of single-channel interactions
between the particles. These would describe the so-called open-channel-dominated
resonances in real systems. A more general description of resonances
would require at least two channels.

\noindent %
\noindent\begin{minipage}[t]{1\columnwidth}%
\rule[0.5ex]{1\columnwidth}{1pt}%
\end{minipage}

\subsection*{Acknowledgments}

C. S. and P. N. acknowledge support from the JSPS Grants-in-Aid for
Scientific Research on Innovative Areas (No. JP18H05407).
\hypersetup{urlcolor=myred}

\noindent {\small{}{} \bibliographystyle{IEEEtran2}
\bibliography{paper_Ludo4}
 }{\small\par}

\clearpage{}

\section*{Appendix 1: separable potential for different partial waves}

A separable potential acting on the $\ell$th partial wave has the
general form: 
\begin{equation}
\hat{V}=\sum_{m=-\ell}^{\ell}\lambda_{\ell,m}\vert\Phi_{\ell,m}\rangle\langle\Phi_{\ell,m}\vert,\label{eq:SeparablePotential}
\end{equation}
with 
\begin{equation}
\langle\bm{k}\vert\Phi_{\ell,m}\rangle=\Phi_{\ell,m}(\bm{k})=k^{\ell}\phi_{\ell,m}(k)\sqrt{4\pi}Y_{\ell,m}(\hat{\bm{k}})\label{eq:Chi}
\end{equation}
where $Y_{\ell,m}$ are the spherical harmonics and $\hat{\bm{k}}$
is the unit vector $\bm{k}/k$.

The action of this separable potential on a wave function $\Psi$
is given in momentum representation by $\langle\bm{k}\vert\hat{V}\vert\Psi\rangle$,
i.e. 
\begin{align}
 & \sum_{m=-\ell}^{\ell}\lambda_{\ell m}\int d^{3}\bm{k}^{\prime}\Phi_{\ell,m}^{*}(\bm{k}^{\prime})\Phi_{\ell,m}(\bm{k})\langle\bm{k}^{\prime}\vert\Psi\rangle\nonumber \\
 & =\sum_{m=-\ell}^{\ell}4\pi\lambda_{\ell,m}k^{\ell}\phi_{\ell,m}(k)Y_{\ell,m}(\hat{\bm{k}})\int_{0}^{\infty}dk^{\prime}k^{\prime\ell}\phi_{\ell,m}^{*}(k^{\prime})\nonumber \\
 & \qquad\times\underbrace{\int d^{2}\hat{\bm{k}}^{\prime}Y_{\ell,m}^{*}(\hat{\bm{k}}^{\prime})\langle\bm{k}^{\prime}\vert\Psi\rangle}_{\Psi_{\ell,m}(k^{\prime})}
\end{align}
where $\Psi_{\ell,m}$ is the partial wave ($\ell,m$) of the wave
function $\Psi$. Thus we see that the potential only acts on that
partial wave.

For the $s$ wave ($\ell=0$) we have $Y_{0,0}(\hat{\bm{k}})=1/\sqrt{4\pi}$,
and thus, 
\begin{equation}
\hat{V}=\lambda_{0,0}\vert\phi_{0,0}\rangle\langle\phi_{0,0}\vert,
\end{equation}
which is the form of Eq.~(\ref{eq:V}) with $\lambda_{0,0}\equiv\frac{2\pi\hbar^{2}}{\mu_{23}}\xi$
and $\phi_{0,0}\equiv\chi$.

For the $p$ wave ($\ell=1$) we have $Y_{1,m}(\hat{\bm{k}})=\sqrt{\frac{3}{4\pi}}\hat{\bm{k}}\cdot\hat{\bm{e}}_{m}$,
and thus, 
\begin{align}
\hat{V} & =\sum_{m=-1}^{1}\lambda_{1,m}\vert\Phi_{1,m}\rangle\langle\Phi_{1,m}\vert,
\end{align}
with 
\begin{equation}
\Phi_{1,m}(\bm{k})=k\phi_{1,m}(k)\sqrt{3}\hat{\bm{k}}\cdot\hat{\bm{e}}_{m},
\end{equation}
which is the form of Eq.~(\ref{eq:U}), with $\lambda_{1,m}\equiv\frac{2\pi\hbar^{2}}{\mu_{12}}g_{m}$
and $\Phi_{1,m}(\bm{k})\equiv\Phi_{m}(\bm{k})$, i.e. $\phi_{1,m}(k)\equiv\phi_{m}(k)/\sqrt{3}$.

The Tables (\ref{tab:S-wave-Potentials}) and (\ref{tab:P-wave-Potentials})
give the different parameters of the potentials used in this paper.

\begin{table*}
\begin{tabular}{|c|c|c|c|}
\hline 
s-wave model  & $\chi(k)$  & $\vert\chi(i\kappa)\vert^{2}/f_{0}(i\kappa)$  & $\rs$ \tabularnewline
\hline 
\hline 
Gaussian  & $\exp(-k^{2}/\Lambda_{0}^{2})$  & $\frac{1}{a}-\kappa e^{\frac{2\kappa^{2}}{\Lambda_{0}^{2}}}\text{erfc}\left(\frac{\sqrt{2}\kappa}{\Lambda_{0}}\right)$  & $\frac{4\sqrt{\frac{2}{\pi}}}{\Lambda_{0}}$\tabularnewline
\hline 
Yamaguchi  & $\frac{1}{1+k^{2}/\Lambda_{0}^{2}}$  & $\frac{1}{a}-\frac{\kappa\Lambda_{0}(\kappa+2\Lambda_{0})}{2(\kappa+\Lambda_{0})^{2}}$  & $\frac{3}{\Lambda_{0}}$\tabularnewline
\hline 
Yamaguchi-squared  & $\left[\frac{1}{1+k^{2}/\Lambda_{0}^{2}}\right]^{2}$  & $\frac{1}{a}-\frac{\kappa\Lambda_{0}\left(5\kappa^{3}+20\kappa^{2}\Lambda_{0}+29\kappa\Lambda_{0}^{2}+16\Lambda_{0}^{3}\right)}{16(\kappa+\Lambda_{0})^{4}}$  & $\frac{35}{8\Lambda_{0}}$\tabularnewline
\hline 
Cut-off  & $\begin{cases}
1 & k\le\Lambda_{0}\\
0 & k>\Lambda_{0}
\end{cases}$  & $\frac{1}{a}-\frac{2\kappa\tan^{-1}\left(\frac{\Lambda_{0}}{\kappa}\right)}{\pi}$  & $\frac{4}{\pi\Lambda_{0}}$\tabularnewline
\hline 
\end{tabular}

\caption{\label{tab:S-wave-Potentials}Separable potential models used for
the s-wave interaction $V$ between a fermion and the particle. For
each model, the table provides the explicit expression of the form
factor $\chi$, the s-wave scattering amplitude $f_{0}$, and the
parameter $\rs$ corresponding to the s-wave effective range at resonance.}
\end{table*}

\begin{table*}
\begin{tabular}{|c|c|c|c|}
\hline 
p-wave model  & $\phi(k)$  & $\vert\kappa\phi(i\kappa)\vert^{2}/f_{1}(i\kappa)$  & $\alphas$ \tabularnewline
\hline 
\hline 
Gaussian  & $\exp(-k^{2}/\Lambda_{1}^{2})$  & $\frac{1}{v}-\frac{\kappa^{2}\Lambda_{1}}{\sqrt{2\pi}}+\kappa^{3}e^{\frac{2\kappa^{2}}{\Lambda_{1}^{2}}}\text{erfc}\left(\frac{\sqrt{2}\kappa}{\Lambda_{1}}\right)$  & $\frac{\Lambda_{1}}{\sqrt{2\pi}}$\tabularnewline
\hline 
Yamaguchi  & $\frac{1}{1+k^{2}/\Lambda_{1}^{2}}$  & $\frac{1}{v}-\frac{\kappa^{2}\Lambda_{1}^{3}}{2(\kappa+\Lambda_{1})^{2}}$  & $\frac{\Lambda_{1}}{2}$\tabularnewline
\hline 
Yamaguchi squared  & $\left[\frac{1}{1+k^{2}/\Lambda_{1}^{2}}\right]^{2}$  & $\frac{1}{v}-\frac{\kappa^{2}\Lambda_{1}^{3}\left(\kappa^{2}+4\kappa\Lambda_{1}+5\Lambda_{1}^{2}\right)}{16(\kappa+\Lambda_{1})^{4}}$  & $\frac{5\Lambda_{1}}{16}$\tabularnewline
\hline 
Cut-off  & $\begin{cases}
1 & k\le\Lambda_{1}\\
0 & k>\Lambda_{1}
\end{cases}$  & $\frac{1}{v}-\frac{2}{\pi}\kappa^{2}\left(\Lambda_{1}-\kappa\tan^{-1}\left(\frac{\Lambda_{1}}{\kappa}\right)\right)$  & $\frac{2}{\pi}\Lambda_{1}$\tabularnewline
\hline 
\end{tabular}

\caption{\label{tab:P-wave-Potentials}Separable potential models used for
the p-wave interaction $U$ between the two fermions. For each model,
the table provides the explicit expression of the form factor $\phi$,
the p-wave scattering amplitude $f_{1}$, and the \alphasname $\alphas$
corresponding to the inverse of the effective range at the p-wave
resonance.}
\end{table*}

\section*{Appendix 2: derivation of the STM equations}

Here we derive the STM equations~(\ref{eq:STM1}-\ref{eq:STM2}).

First, we use the separable forms~(\ref{eq:V}) and~(\ref{eq:U})
of the interactions $\hat{V}$ and $\hat{U}$ in the Schrödinger equation~(\ref{eq:SchrodingerEq}).
This gives:
\begin{align}
\left(\frac{1}{2}k_{1}^{2}+\frac{1}{2}k_{2}^{2}+\frac{x}{2}k_{3}^{2}+q^{2}\right)\langle\{\bm{k}_{i}\}\vert\Psi\rangle\nonumber \\
-12\pi\sum_{m=-1}^{1}\langle\bm{k}_{12}\vert\Phi_{m}\rangle P_{m}(\bm{k}_{3})\nonumber \\
-4\pi y\langle\bm{k}_{23}\vert\chi\rangle S(\bm{k}_{1})+4\pi y\langle\bm{k}_{31}\vert\chi\rangle S^{\prime}(\bm{k}_{2}) & =0,\label{eq:SchrodingerEq2}
\end{align}
where $x=M/m$ is the mass ratio, and $y$ is the short-hand notation
for $(1+x)/2$. The relative wave functions $S$ and $P_{m}$ are
defined by Eqs.~(\ref{eq:S}-\ref{eq:P}), and $S^{\prime}$ is given
by:
\begin{equation}
S^{\prime}(\bm{k}_{2})=-\xi\int\frac{d^{3}k_{31}}{(2\pi)^{3}}\langle\chi\vert\bm{k}_{31}\rangle\langle\{\bm{k}_{i}\}\vert\Psi\rangle.\label{eq:S'}
\end{equation}
From the antisymmetry of the wave function under the exchange of
the fermionic particles 1 and 2, i.e. $\langle\bm{k}_{2},\bm{k}_{1},\bm{k}_{3}\vert\Psi\rangle=-\langle\bm{k}_{1},\bm{k}_{2},\bm{k}_{3}\vert\Psi\rangle$,
it is easy to check from Eq.~(\ref{eq:S'}) that $S^{\prime}(\bm{k})=-S(\bm{k})$.
The three-body wave function $\langle\{\bm{k}_{i}\}\vert\Psi\rangle/4\pi$
is thus given by:

\begin{equation}
\frac{3\sum_{m}\!\langle\bm{k}_{12}\vert\Phi_{m}\rangle P_{m}(\bm{k}_{3})\!+\!y\!\left(\langle\bm{k}_{23}\vert\chi\rangle S(\bm{k}_{1})\!-\!\langle\bm{k}_{31}\vert\chi\rangle S(\bm{k}_{2})\right)}{\frac{1}{2}k_{1}^{2}+\frac{1}{2}k_{2}^{2}+\frac{x}{2}k_{3}^{2}+q^{2}}.\label{eq:WaveFunction}
\end{equation}

Inserting this expression in the definitions of $S$ and $P_{m}$,
Eqs.~(\ref{eq:S}-\ref{eq:P}), one obtains:
\begin{align}
 & -\frac{S(\bm{k}_{1})}{4\pi\xi}=3\!\!\int\!\frac{d^{3}k_{23}}{(2\pi)^{3}}\langle\chi\vert\bm{k}_{23}\rangle\frac{\sum_{m}\!\langle\bm{k}_{12}\vert\Phi_{m}\rangle P_{m}(\bm{k}_{3})}{\frac{1}{2}k_{1}^{2}+\frac{1}{2}k_{2}^{2}+\frac{x}{2}k_{3}^{2}+q^{2}}\nonumber \\
 & \qquad+y\!\!\int\!\frac{d^{3}k_{23}}{(2\pi)^{3}}\langle\chi\vert\bm{k}_{23}\rangle\frac{\langle\bm{k}_{23}\vert\chi\rangle S(\bm{k}_{1})\!-\!\langle\bm{k}_{31}\vert\chi\rangle S(\bm{k}_{2})}{\frac{1}{2}k_{1}^{2}+\frac{1}{2}k_{2}^{2}+\frac{x}{2}k_{3}^{2}+q^{2}},\label{eq:STM1a}\\
 & -\frac{P_{m}(\bm{k}_{3})}{4\pi g_{m}}=3\!\!\int\!\frac{d^{3}k_{12}}{(2\pi)^{3}}\langle\Phi_{m}\vert\bm{k}_{12}\rangle\frac{\sum_{m^{\prime}}\!\langle\bm{k}_{12}\vert\Phi_{m^{\prime}}\rangle P_{m^{\prime}}(\bm{k}_{3})}{\frac{1}{2}k_{1}^{2}+\frac{1}{2}k_{2}^{2}+\frac{x}{2}k_{3}^{2}+q^{2}}\nonumber \\
 & \qquad+y\!\!\int\!\frac{d^{3}k_{12}}{(2\pi)^{3}}\langle\Phi_{m}\vert\bm{k}_{12}\rangle\frac{\langle\bm{k}_{23}\vert\chi\rangle S(\bm{k}_{1})\!-\!\langle\bm{k}_{31}\vert\chi\rangle S(\bm{k}_{2})}{\frac{1}{2}k_{1}^{2}+\frac{1}{2}k_{2}^{2}+\frac{x}{2}k_{3}^{2}+q^{2}}.\label{eq:STM2a}
\end{align}
From the translational invariance of the system, the total momentum
can be set to zero:
\begin{equation}
\bm{k}_{1}+\bm{k}_{2}+\bm{k}_{3}=0.\label{eq:TotalMomentumZero}
\end{equation}
It follows that the three vectors $\bm{k}_{1}$, $\bm{k}_{2}$, and
$\bm{k}_{3}$ can be expressed in terms of two Jacobi vectors $\bm{k}_{ij}$
and $\bm{k}_{k}$:
\begin{align}
\bm{k}_{1}= & \bm{k}_{31}-\frac{x}{1+x}\bm{k}_{2}=-\bm{k}_{12}-\frac{1}{2}\bm{k}_{3},\label{eq:k1expr}\\
\bm{k}_{2}= & -\bm{k}_{23}-\frac{x}{1+x}\bm{k}_{1}=\bm{k}_{12}-\frac{1}{2}\bm{k}_{3},\label{eq:k2expr}\\
\bm{k}_{3}= & +\bm{k}_{23}-\frac{1}{1+x}\bm{k}_{1}=-\bm{k}_{31}-\frac{1}{1+x}\bm{k}_{2}.\label{eq:k3expr}
\end{align}
In particular, the denominator in Eqs.~(\ref{eq:STM1a}-\ref{eq:STM2a})
admits the following expressions:
\begin{align}
\frac{1}{2}k_{1}^{2}+\frac{1}{2}k_{2}^{2}+\frac{x}{2}k_{3}^{2}+q^{2} & =yk_{23}^{2}+k_{1}^{2}\frac{2x+1}{4y}+q^{2}\label{eq:denominator1}\\
 & =k_{12}^{2}+\frac{2x+1}{4}k_{3}^{2}+q^{2}\label{eq:denominator2}
\end{align}

Now, one can see that the term proportional to $S(\bm{k}_{1})$ in
the right-hand side of Eq.~(\ref{eq:STM1a}) can be factored with
the left-hand side. Similarly, the term proportional to $P_{m^{\prime}}(\bm{k}_{3})$
in the right-hand side of Eq.~(\ref{eq:STM2a}) can be shown to be
proportional to $P_{m}(\bm{k}_{3})$, by performing the integration
over the orientation of $\bm{k}_{12}$ and using the relation $\int d^{2}\hat{\bm{k}}_{12}\left(\hat{\bm{k}}_{12}\cdot\hat{\bm{e}}_{m}\right)\left(\hat{\bm{k}}_{12}\cdot\hat{\bm{e}}_{m^{\prime}}\right)=\frac{4\pi}{3}\delta_{m,m^{\prime}}$.
It can therefore be factored with the left-hand side. Using Eq.~(\ref{eq:denominator1})
in Eq.~(\ref{eq:STM1a}) and Eq.~(\ref{eq:denominator2}) in Eq.~(\ref{eq:STM2a}),
one obtains:
\begin{align}
 & \frac{S(\bm{k}_{1})}{4\pi\Xi_{0,0}}=-3\!\!\int\!\frac{d^{3}k_{23}}{(2\pi)^{3}}\langle\chi\vert\bm{k}_{23}\rangle\frac{\sum_{m}\!\langle\bm{k}_{12}\vert\Phi_{m}\rangle P_{m}(\bm{k}_{3})}{\frac{1}{2}k_{1}^{2}+\frac{1}{2}k_{2}^{2}+\frac{x}{2}k_{3}^{2}+q^{2}}\nonumber \\
 & \qquad\qquad+y\!\!\int\!\frac{d^{3}k_{23}}{(2\pi)^{3}}\langle\chi\vert\bm{k}_{23}\rangle\frac{\!\langle\bm{k}_{31}\vert\chi\rangle S(\bm{k}_{2})}{\frac{1}{2}k_{1}^{2}+\frac{1}{2}k_{2}^{2}+\frac{x}{2}k_{3}^{2}+q^{2}},\label{eq:STM1b}\\
 & \frac{P_{m}(\bm{k}_{3})}{4\pi\Xi_{1,m}}=y\!\!\int\!\frac{d^{3}k_{12}}{(2\pi)^{3}}\langle\Phi_{m}\vert\bm{k}_{12}\rangle\nonumber \\
 & \qquad\qquad\qquad\times\frac{\langle\bm{k}_{31}\vert\chi\rangle S(\bm{k}_{2})\!-\!\langle\bm{k}_{23}\vert\chi\rangle S(\bm{k}_{1})}{\frac{1}{2}k_{1}^{2}+\frac{1}{2}k_{2}^{2}+\frac{x}{2}k_{3}^{2}+q^{2}}.\label{eq:STM2b}
\end{align}
with
\begin{align}
\frac{1}{\Xi_{0,0}} & =\frac{1}{\xi}+4\pi y\int\frac{d^{3}k_{23}}{(2\pi)^{3}}\frac{\vert\chi(k_{23})\vert^{2}}{yk_{23}^{2}+k_{1}^{2}\frac{2x+1}{4y}+q^{2}},\label{eq:F00}\\
\frac{1}{\Xi_{1,m}} & =\frac{1}{g_{m}}+4\pi\int\frac{d^{3}k_{12}}{(2\pi)^{3}}\frac{\vert\phi_{m}(k_{12})\vert^{2}k_{12}^{2}}{k_{12}^{2}+\frac{2x+1}{4}k_{3}^{2}+q^{2}},\label{eq:F1m}
\end{align}
which can be cast in the form of the right-hand sides of Eqs.~(\ref{eq:f0}-\ref{eq:f1})
by relabelling the variables $k_{23}$ and $k_{12}$ to $k$ and introducing
the relative momenta $\kappa$ given by Eqs.~(\ref{eq:kappas}-\ref{eq:kappap}).

Finally, using Eqs\@.~(\ref{eq:k1expr}-\ref{eq:k3expr}), one can
make a change of the integration variables in Eq.~(\ref{eq:STM1b}-\ref{eq:STM2b})
such that they correspond to the variables of the functions $P_{m}$
and $S$. This change shows that the terms proportional to $S(\bm{k}_{2})$
and $S(\bm{k}_{1})$ in Eq.~(\ref{eq:STM2b}) give the same contribution,
which leads to the equations,
\begin{align}
 & \frac{S(\bm{k}_{1})}{4\pi\Xi_{0,0}}=y\!\!\int\!\frac{d^{3}k_{2}}{(2\pi)^{3}}\frac{\langle\chi\vert\bm{k}_{2}+\frac{x}{2y}\bm{k}_{1}\rangle\!\langle\bm{k}_{1}+\frac{x}{2y}\bm{k}_{2}\vert\chi\rangle S(\bm{k}_{2})}{y\left(k_{1}^{2}+k_{2}^{2}\right)+x\bm{k}_{1}\cdot\bm{k}_{2}+q^{2}}\label{eq:STM1c}\\
 & \qquad+3\!\!\int\!\frac{d^{3}k_{3}}{(2\pi)^{3}}\frac{\langle\chi\vert\bm{k}_{3}+\frac{\bm{k}_{1}}{2y}\rangle\sum_{m}\!\langle\bm{k}_{1}+\frac{\bm{k}_{3}}{2}\vert\Phi_{m}\rangle P_{m}(\bm{k}_{3})}{k_{1}^{2}+yk_{3}^{2}+\bm{k}_{3}\cdot\bm{k}_{1}+q^{2}},\nonumber \\
 & \frac{P_{m}(\bm{k}_{3})}{4\pi\Xi_{1,m}}=2y\!\!\int\!\frac{d^{3}k_{1}}{(2\pi)^{3}}\frac{\langle\Phi_{m}\vert\bm{k}_{1}+\frac{\bm{k}_{3}}{2}\rangle\langle\bm{k}_{3}+\frac{\bm{k}_{1}}{2y}\vert\chi\rangle S(\bm{k}_{1})}{k_{1}^{2}+yk_{3}^{2}+\bm{k}_{1}\cdot\bm{k}_{3}+q^{2}}.\label{eq:STM2c}
\end{align}
Relabelling the non-integrated vector as $\bm{k}$ and the integrated
vector as $\bm{k}^{\prime}$ in both equations, one arrives at the
STM equations~(\ref{eq:STM1}-\ref{eq:STM2}). The quantities $\Xi_{0,0}$
and $\Xi_{1,m}$ are directly related to the two-body scattering amplitudes
of the potentials $\hat{V}$ and $\hat{U}$ given respectively by:
\begin{align}
\frac{1}{\Xi_{0,0}} & =-\frac{\vert\chi(i\kappa_{0})\vert^{2}}{f_{0,0}(i\kappa_{0})}\label{eq:Xi00}\\
\frac{1}{\Xi_{1,m}} & =\frac{\vert\kappa\phi_{m}(i\kappa_{1})\vert^{2}}{f_{1,m}(i\kappa_{1})}\label{eq:Xi1m}
\end{align}
where $\kappa_{0}$ and $\kappa_{1}$ are defined in Eqs.~(\ref{eq:kappas}-\ref{eq:kappap}).

\section*{Appendix 3: form of the functions $S$ and $\bm{P}$}

Here we derive the forms of $S$ and $\bm{P}$ functions for a three-body
state of total angular momentum and parity symmetry $J^{\pi}=1^{-}$.
The same considerations of symmetry were done in Ref.~\cite{Jona-Lasinio2008}
in the case of three identical fermions.

\subsection*{Form of $S$}

$S$ depends on the momentum $\bm{k}$ between the fermion-particle
subsystem (13) and fermion 2. Since the fermion-particle susbsystem
is assumed to have an angular momentum $l=0$ and the total angular
momentum is $J=1$, the only possible angular momentum $L$ between
(13) and 2 is 1. The function $S$ is thus proportional to an angular
momentum state $L=1$, with a proportionality factor that depends
on the norm $k$ of $\bm{k}$.

\begin{align}
S(\bm{K}) & \propto\vert\stackrel{J}{1},\stackrel{M_{J}}{0}\rangle_{0\oplus1}\nonumber \\
 & \propto\vert\stackrel{l}{0},\stackrel{m}{0}\rangle\vert\stackrel{L}{1},\stackrel{M}{0}\rangle\nonumber \\
 & =s(k)(\bm{e}_{k}\cdot\bm{e}_{z}),\label{eq:formOfS}
\end{align}
which is Eq.~(\ref{eq:Sform}). Here, we use the fact that the angular
momentum $\vert\stackrel{L}{1},\stackrel{M}{0}\rangle$ is proportional
to $\bm{e}_{k}\cdot\bm{e}_{z}$, where $\bm{e}_{k}=\bm{k}/k$ is the
unit vector along $\bm{k}$ and $\bm{e}_{z}$ is a fixed unit vector
along which the projection of angular momentum is assumed to be zero.

\subsection*{Form of $\bm{P}$}

$\bm{P}$ depends on the momentum $\bm{k}$ between the fermion-fermion
subsystem (12) and third particle 3. Since the fermion-fermion subsystem
is assumed to have an angular momentum $l=1$ and the total angular
momentum is $J=1$, the only possible angular momentum $L$ between
(12) and 3 is either 0 or 2. From the negative parity, we further
restrict to the two angular momentum compositions: 
\begin{equation}
\vert\stackrel{l}{1},\stackrel{m}{0}\rangle\vert\stackrel{L}{0},\stackrel{M}{0}\rangle
\end{equation}
and 
\begin{multline}
\sqrt{\frac{3}{10}}\vert\stackrel{l}{1},\stackrel{m}{-1}\rangle\vert\stackrel{L}{2},\stackrel{M}{1}\rangle\\
-\sqrt{\frac{2}{5}}\vert\stackrel{l}{1},\stackrel{m}{0}\rangle\vert\stackrel{L}{2},\stackrel{M}{0}\rangle+\sqrt{\frac{3}{10}}\vert\stackrel{l}{1},\stackrel{m}{1}\rangle\vert\stackrel{L}{2},\stackrel{M}{-1}\rangle.
\end{multline}
The quantity $\bm{P}(\bm{k})\cdot\bm{k}_{12}$ is therefore a linear
combination of the above two angular momentum states, where the linear
coefficients depend on the norm $K$. Expressing these angular momentum
states in terms of scalar and vector products, one finds: 
\begin{equation}
\bm{P}(\bm{k})=p_{0}(k)\bm{e}_{z}+p_{2}(k)\left[\bm{e}_{k}\times(\bm{e}_{k}\times\bm{e}_{z})\right],
\end{equation}
which is Eq.~(\ref{eq:Pform}). 

\section*{Appendix 4: perturbation of STM equations}

We write the STM equation at negative energy in the generic form 
\begin{equation}
\int\frac{d^{3}k'}{(2\pi)^{3}}\langle\mathbf{k}|\mathcal{M}(q)|\mathbf{k}'\rangle\langle\mathbf{k}'|\lambda\rangle=0
\end{equation}
where ${\mathcal{M}(q)}$ may be a matrix as in Eqs.~(\ref{eq:STM1},~\ref{eq:STM2})
or a scalar operator, as it is the case for example when there is
no p-wave interaction. We divide the STM operator into two terms:
${\mathcal{M}(q)=\mathcal{M}^{(0)}(q)+\mathcal{M}^{(1)}(q)}$ and
consider the situation where ${\mathcal{M}^{(1)}(q)}$ can be treated
as a perturbation with respect to the dominant term ${\mathcal{M}^{(0)}(q)}$.
At the first order of perturbation, we decompose the eigenvector and
the binding wavenumber as ${|\lambda\rangle=|\lambda^{(0)}\rangle+|\lambda^{(1)}\rangle}$
and ${q=q^{(0)}+q^{(1)}}$, such that at the lowest order, the STM
equation is 
\begin{equation}
\mathcal{M}^{(0)}(q^{(0)})|\lambda^{(0)}\rangle=0.\label{eq:STM_unperturbed}
\end{equation}
At the first order in perturbation, the STM equation is expanded as
\begin{multline}
\mathcal{M}^{(0)}(q^{(0)})|\lambda^{(1)}\rangle+q^{(1)}\frac{\delta\mathcal{M}^{(0)}}{\delta q}\biggr|_{q^{(0)}}|\lambda^{(0)}\rangle\\
+\mathcal{M}^{(1)}(q^{(0)})|\lambda^{(0)}\rangle=0.\label{eq:perturb3}
\end{multline}
Without loss of generality, we assume that ${\langle\mathbf{k}|\mathcal{M}^{(0)}|\mathbf{k}'\rangle}$
and ${\langle\mathbf{k}|\mathcal{M}^{(1)}|\mathbf{k}'\rangle}$ are
written in a symmetric form, so that the perturbed eigenvector verifies
${\langle\lambda^{(0)}|\lambda^{(1)}\rangle=0}$. Applying ${\langle\lambda^{(0)}|}$
on the left of Eq.~\eqref{eq:perturb3}, one obtains 
\begin{equation}
\langle\lambda^{(0)}|\frac{\delta\mathcal{M}^{(0)}}{\delta q}|\lambda^{(0)}\rangle q^{(1)}+\langle\lambda^{(0)}|\mathcal{M}^{(1)}|\lambda^{(0)}\rangle=0
\end{equation}
and thus 
\begin{equation}
q^{(1)}=-\frac{\langle\lambda^{(0)}|\mathcal{M}^{(1)}|\lambda^{(0)}\rangle}{\langle\lambda^{(0)}|\frac{\delta\mathcal{M}^{(0)}}{\delta q}|\lambda^{(0)}\rangle}.\label{eq:pertub1}
\end{equation}
In the particular case where $\mathcal{M}^{(0)}$ is the scalar STM
operator in absence of p-wave interaction, we will show that ${\langle\lambda^{(0)}|\frac{\delta\mathcal{M}^{(0)}}{\delta q}|\lambda^{(0)}\rangle}$
can be expressed in terms of the sweep parameter $\frac{\partial E}{\partial(1/a)}$.
For this purpose, we consider a small variation of the scattering
length in the unperturbed STM equation $a\to a+\delta a$, that induces
a small change of the binding wavenumber $q^{(0)}\to q^{(0)}+\delta q$.
At the first order in $\delta(1/a)$, one has 
\begin{equation}
\langle\lambda^{(0)}|\frac{\delta\mathcal{M}^{(0)}}{\delta q}|\lambda^{(0)}\rangle\delta q+\langle\lambda^{(0)}|\frac{\delta\mathcal{M}^{(0)}}{\delta(1/a)}|\lambda^{(0)}\rangle\delta(1/a)=0.\label{eq:perturb_L0}
\end{equation}
The value of the term ${\langle\lambda^{(0)}|\frac{\delta\mathcal{M}^{(0)}}{\delta(1/a)}|\lambda^{(0)}\rangle}$
depends on a common factor in ${\mathcal{M}}$, ${\mathcal{M}^{(0)}}$,
and ${\mathcal{M}^{(1)}}$. We fix this factor such that $\frac{\delta\mathcal{M}^{(0)}}{\delta(1/a)}=-1$
and thus $\langle\lambda^{(0)}|\frac{\delta\mathcal{M}^{(0)}}{\delta(1/a)}|\lambda^{(0)}\rangle=-\langle\lambda^{(0)}|\lambda^{(0)}\rangle$.
Then, using this normalisation of the STM operator and injecting Eq.~\eqref{eq:perturb_L0}
in Eq.~\eqref{eq:pertub1}, one finds 
\begin{equation}
\boxed{E^{(1)}=-\frac{\partial E^{(0)}}{\partial(1/a)}\frac{\langle\lambda^{(0)}|\mathcal{M}^{(1)}|\lambda^{(0)}\rangle}{\langle\lambda^{(0)}|\lambda^{(0)}\rangle}}\label{eq:perturbation}
\end{equation}
where ${E^{(0)}}$ is the energy of the unperturbed system ${E^{(0)}=-\hbar^{2}q^{(0)}\,^{2}/(2M)}$
and ${E^{(1)}}$ is the shift in energy resulting from the perturbation:
${E^{(1)}=-\hbar^{2}q^{(0)}q^{(1)}/M}$.
\end{document}